\documentclass[12pt]{article}
\usepackage{graphicx}
\usepackage[dvips]{color}
\usepackage{epsfig}

\def\Journal#1#2#3#4{{#1} {\bf #2}, #3 (#4)}
\def\beq{\begin{equation}}
\def\eeq{\end{equation}}

\def\NPB{{\em Nucl. Phys.} B}
\def\PLB{{\em Phys. Lett.}  B}
\def\PRL{\em Phys. Rev. Lett.}
\def\PRD{{\em Phys. Rev.} D}

\def\r2{\sqrt 2}
\def\beq{\begin{equation}}
\def\eeq{\end{equation}}
\def\beqn{\begin{eqnarray}}
\def\eeqn{\end{eqnarray}}

\def\sinW2{\sin^2\theta_W}

\def\mz2{M_{z}^2}
\def\c2b{\cos 2\beta}

\def\m#1{{\tilde m}_#1}

\def\mz{M_z}

\def\Fq2{F_{2}(q^2)}
\def\f{\({\cal F}\)}
\def\d1{{\f(\tilde c;\tilde s;\tilde W)+ \f(\tilde c;\tilde \mu;\tilde W)}}

\def\sec2w{sec^2\theta_W}
\def\mhalf{m_{\frac{1}{2}}}
\def\12{$\frac{1}{2}$}
\def\m0{$m_0$}
\def\bsg{b\rightarrow s \gamma}
\def\bbsg{BR($\bsg$)~}
\begin{document}
\baselineskip 18pt
\def\today{\ifcase\month\or
 January\or February\or March\or April\or May\or June\or
 July\or August\or September\or October\or November\or December\fi
 \space\number\day, \number\year}
\def\thebibliography#1{\section*{References\markboth
 {References}{References}}\list
 {[\arabic{enumi}]}{\settowidth\labelwidth{[#1]}
 \leftmargin\labelwidth
 \advance\leftmargin\labelsep
 \usecounter{enumi}}
 \def\newblock{\hskip .11em plus .33em minus .07em}
 \sloppy
 \sfcode`\.=1000\relax}
\let\endthebibliography=\endlist

\begin{titlepage}

\begin{center}
{\large {\bf WMAP Dark Matter Constraints and Yukawa 
Unification in SUGRA models  with CP  Phases}}\\
\vskip 1.5 true cm
\renewcommand{\thefootnote}
{\fnsymbol{footnote}}
 Mario E. G\'omez$^{a}$,  Tarek Ibrahim$^{b,c}$, Pran Nath$^{c,d}$ and 
 Solveig Skadhauge$^e$  
\vskip 0.8 true cm

\noindent
{\it a. Departamento de F\'{\i}sica Aplicada, Facultad de Ciencias 
Experimentales,}\\
{\it Universidad de Huelva, 21071 Huelva, Spain}\\
{\it b. Department of  Physics, Faculty of Science,
University of Alexandria,} \\
{\it Alexandria, Egypt\footnote{Permanent address of T.I.}}\\ 
{\it c. Department of Physics, Northeastern University,
Boston, MA 02115-5000, USA.} \\
{\it d. Max Planck Institute for Physics, Fohringer Ring 6, D-80805, Munich,
Germany.}\\
{\it e. Instituto de F\'{\i}sica,     
Universidade de S\~{a}o Paulo, 05315-970 S\~{a}o Paulo, SP, Brazil.}

\end{center}
      
\vskip 1.0 true cm
\centerline{\bf Abstract}
\medskip

The compatibility of producing the observed amount of dark matter, 
as indicated by the WMAP data, through the relic abundance of 
neutralinos with Yukawa unification and with the measured 
rate of $b\rightarrow s \gamma$ is analyzed in mSUGRA 
and extended SUGRA unified models with the inclusion of CP phases. 
The CP phases affect the analysis in several ways,  
e.g., through the threshold corrections to the b-quark mass, via their
effects on the neutralino relic density and through the SUSY contribution
 to the BR($b\rightarrow s \gamma$) which is sensitive to the CP phases. 
We present some specific models with large SUSY phases, 
which can accommodate the fermion electric dipole moment 
constraints and give a neutralino relic density in agreement 
with observations as well as  with the b-$\tau$ unification constraint. The possibility of 
achieving WMAP relic density constraints with full Yukawa unification 
is also explored.  

\end{titlepage}
 
\section{Introduction}
The Wilkinson Microwave Anisotropy Probe (WMAP) has placed stringent 
bounds on the amount of cold dark matter (CDM) in the universe.
The amount of CDM  deduced from WMAP data is 
given by\cite{Bennett:2003bz,Spergel:2003cb}
\beqn
\Omega_{CDM} h^2 =0.1126^{+0.008}_{-0.009} \;,
\eeqn
where $\Omega_{CDM} =\rho_{CDM}/\rho_c$, and where $\rho_{CDM}$ is
the matter density of cold dark matter and $\rho_c$ is the 
critical matter density needed to close the universe, and 
$h$ is the Hubble parameter measured in units of 100km/s/Mpc.
It is reasonable to assume that similar amounts of dark matter
exist in our Milky Way and in the terrestrial neighborhood, and there
are many ongoing experiments for the detection of such dark matter
in the laboratory.
On the theoretical side the WMAP data on cold dark matter puts 
stringent constraints on unified models of fundamental interactions
since such models are called upon to predict or at least accommodate
the WMAP data on CDM. As is well known, supergravity 
unified models~\cite{msugra} with R-parity conservation 
allow for the possibility that the lightest neutralino may be 
the lightest supersymmetric particle (LSP) which could serve as a 
dark matter candidate~\cite{goldberg}\footnote{There is also revived 
interest in the possibility that the LSP in SUGRA models could be 
the gravitino. For an update see Ref.\cite{Ellis:2003dn}}. 
A hallmark of many unified models is Yukawa unification. 
In this paper we carry out a detailed investigation of the possibility
of accommodating the WMAP cold dark matter data in the neutralino 
LSP scenario but under the constraints of Yukawa unification and 
including the effect of CP phases\footnote{For an analysis of dark 
matter with CP phases but without inclusion of the Yukawa unification 
constraints see Refs.\cite{Gomez:2004ek,cin}. An analysis of dark 
matter with quasi Yukawa unification was given in Refs.\cite{GLP1,GLP2}.}.
Another important constraint is the FCNC constraint given by  
$b\rightarrow s+\gamma$ which is discussed in some detail 
in this paper and included in the analysis.

Since the main focus of the analysis is the Yukawa unification 
constraint on dark matter in SUGRA models\footnote{For recent 
works on dark matter analyses in SUGRA models see
Ref.\cite{sugradark}}, we briefly discuss some broad features 
of this constraint with details to follow later. In the 
supersymmetric framework the unification of the Yukawa couplings 
of the third generation, as predicted in several grand unification 
models, is rather sensitive to the parameters of the SUSY models.
Thus, the compatibility of $b-\tau$ unification at the grand 
unification scale with the observed $b$ and $\tau$ masses depends 
sensitively on the sign of  $\mu$\footnote{We use 
the sign convention on $\mu$ as in Ref.\cite{sugraworking}.} 
(where $\mu$ is the Higgs mixing parameter) as well as on the details 
of the sparticle spectrum~\cite{deBoer:2001xp,Komine:2001rm}.
Moreover, for most of the available parameter space 
b-$\tau$ unification is in conflict with other experimental 
constraints such as the FCNC process $b\rightarrow s \gamma$. 
The more stringent $b-t-\tau$ unification is predicted in the 
minimal $SO(10)$ models where the quarks and leptons, residing 
in the 16-plet spinor representation of $SO(10)$, gain masses 
via coupling with a 10-plet tensor representation of 
$SO(10)$\footnote{More Higgs multiplets are needed to break the gauge 
symmetry correctly down to the standard model gauge symmetry, 
but typically these additional Higgs fields do 
not have couplings to quarks and leptons.}
~\cite{shafi,Hall:1993gn,Baer:1999mc,Baer:2000jj}.
Finally, we mention that there can be GUT scale threshold corrections 
to the Yukawa unification. However, typically they are expected to be 
small~\cite{Wright:1994qb}.

Let us now be more specific and review the situation of Yukawa 
coupling unification in the mSUGRA case with no phases. 
With universal Yukawa couplings at the grand unification 
scale, the masses of the bottom and the top quark are naturally 
higher than the $\tau$ lepton mass. This phenomenon arises 
because of the color interactions which causes the Yukawa 
couplings of the quarks to increase as one goes down to 
lower energy scales. Thus, the running quark masses end up larger 
than the running charged lepton masses. To convert the running mass 
to the pole mass, one needs to include the supersymmetric 
as well as the standard model (SM) threshold corrections. 
In particular, it is well known that the supersymmetric 
threshold correction to the bottom quark mass, $\Delta m_b$, 
can be very large. The value of $\Delta m_b$ is enhanced 
for large values of $\tan\beta$, where $\tan\beta$ 
is the ratio $\langle H_u\rangle / \langle H_d \rangle$ 
and where $H_u$ gives mass to the up 
quark and $H_d$ gives mass to the down quark and the lepton.
In the mSUGRA case with no CP phases, $\Delta m_b$ takes the sign 
of $\mu$ \cite{Hall:1993gn} (unless the trilinear terms are very 
large). A negative SUSY threshold correction to $m_b$ is 
required in models with $b-\tau$ unification, in order to 
obtain a b-quark mass in the allowed range. Therefore, b-$\tau$ 
unification points toward a negative value of the $\mu$-parameter. 
 However, a negative $\mu$ parameter makes the SUSY 
contribution to BR($b\rightarrow s \gamma$) positive and 
hence it adds to the SM contribution and the charged Higgs 
contribution. As a result, a heavy spectrum is required in 
order not to exceed the upper bound for this branching ratio. 

The SUSY contribution to the muon anomalous magnetic moment
also takes the sign of $\mu$ in mSUGRA~\cite{lncn}, and more 
generally this contribution depends on CP phases~\cite{ing2}. 
However, experimentally the situation is less clear 
regarding the implications of the $g_{\mu}-2$ data. 
Thus, while the BNL experiment has significantly
improved the accuracy of the $g_{\mu}-2$ measurement 
\cite{Bennett:2004pv}, ambiguities in the hadronic error, 
which is needed to compute the deviation of the observed 
value from the Standard Model prediction, still persist. 
Currently, the largest source of error in the computation 
of the Standard Model prediction is the O($\alpha^2$) 
hadronic vacuum polarization correction. The most recent 
evaluation of this correction are done by (i) Davier 
et.al.\cite{Davier:2003pw} using the $\tau$ decay data, 
and by (ii) Hagiwara et.al.\cite{Hagiwara:2003da} 
using the low energy data from $e^+e^-\rightarrow$ hadrons.
Assuming that the entire difference $\Delta a_{\mu}$ (where 
$a_{\mu}$ is defined so that the effective operator is 
$a_{\mu}(e/2m_{\mu}) \bar\mu\sigma_{\alpha\beta} \mu F^{\alpha\beta}$)
between experiment and theory comes from supersymmetry, 
one finds that the supersymmetric contribution for the case 
of Davier et.al. is $(7.6\pm 9.0)\times 10^{-10}$, while for 
the case of Hagiwara et.al., the value is 
$(23.9\pm 10.0)\times 10^{-10}$. 
In this analysis we adopt solution (i). In fact, in most of 
the parameter space we explore the sparticle spectrum is rather 
heavy, and the SUSY contribution $\Delta a_{\mu}$ small, and 
thus the $a_{\mu}$ prediction is essentially the same as the
Standard Model prediction which is consistent with the current data.
Solution (ii) puts a more stringent constraint. 
However, the constraint can be softened if 
the universality condition  on the soft terms is 
removed~\cite{Chattopadhyay:2001va,Pallis:2003aw,Profumo:2003sx}.

As indicated earlier this paper is devoted mostly to an 
analysis of the WMAP data with the $b-\tau$ unification
constraint. However, we will also briefly discuss  b-$\tau$-t 
unification. As is well known such a unification requires 
large $\tan\beta$ and for this reason much of the parameter 
space is excluded since it does not correctly break the 
electro-weak symmetry. Several studies has been done, and 
models such as, e.g., the D-term splitting in SO(10) or 
non-universal Higgs masses can indeed give rise to a viable 
t-b-$\tau$ unification~
\cite{Dermisek:2003vn,Balazs:2003mm,Auto:2003ys,Baer:2004xx,Auto:2004km}. 
However, typically these solutions require a very heavy 
SUSY spectrum. Thus, the predicted dark matter abundance of 
neutralinos, in models with R-parity conservation, will be too high 
and thus will over-close the universe. As mentioned earlier, models 
with quasi unification have also been investigated~\cite{GLP1,GLP2}. 

In the present work we first analyze the relic density within 
mSUGRA and show that there exist regions of the parameter 
space where the  WMAP relic density constraint, the Yukawa 
unification constraint, and the \bbsg constraint can all be 
simultaneously satisfied. We then extend the mSUGRA parameter 
space retaining universality on the magnitude of the soft parameters 
but allowing non-universality for the phases in some sectors.
The SUSY contribution to $\Delta m_b$ is phase--dependent 
\cite{Ibrahim:2003ca,Gomez:2004ek} and this allows one 
to determine the phases in some cases such as to obtain $m_b(M_Z)$ 
in the experimental range and thus achieve b-$\tau$ unification. 
Indeed, one finds that with the inclusion of phases $b-\tau$ 
unification is achievable in a large area of the parameter-space. 
Further, it is possible to find arrangement of phases such 
that the prediction of the electric dipole moments (EDMs) 
is in agreement with the experimental bounds.  The case of 
full Yukawa unification is, however, still (almost) incompatible 
with the experimental value for \bbsg. However, it is worth 
keeping in mind that small flavor mixings in the sfermion 
mass matrices can substantially change the predictions 
for \bbsg while leaving other predictions essentially 
untouched. Thus, in principle, non-zero CP-phases could
also allow viable b-$\tau$-t unification modulo mixings 
in the squark flavor sector.
However, we do not pursue this line of investigation in the work here.

The outline of the rest of the paper is as follows: In Sec.(2) 
we give a discussion of the parameter space of the model 
and the details of the procedure of the calculations.  
In Sec.(3) we discuss the calculation of the \bbsg ~and resolve 
some of the ambiguities present in the literature 
in the large $\tan\beta$ enhanced contributions, 
by carrying out an independent analysis of the parameters 
$\epsilon_b^{'}(t), \epsilon_t^{'}(s), \epsilon_{bb}$, which 
codify these contributions. 
In Sec.(4) we carry out an analysis of the relic 
density with the $b-\tau$ unification constraint, and 
with the \bbsg ~constraint both within mSUGRA and in 
extended SUGRA models with phases. In Sec.(5) we give an
analysis of the relic density for the case of the full 
Yukawa unification. In Sec.(6) we discuss the consistency 
of the analysis with large phases with the EDM constraints and 
give examples of models with large phases consistent with the 
WMAP data, b-$\tau$ unification and with the EDM constraints. 
Conclusions are given in Sec.(7).

\section{Constraints on SUGRA models with CP-phases}
Within mSUGRA there are only two physical phases, 
which can be chosen as $\theta_\mu$, the phase of 
the Higgs mixing parameter $\mu$, and $\alpha_0$ 
the phase of the universal trilinear term $A_0$. 
These phases are severely constrained
by the non-observation of the electric dipole moments (EDM). 
The present upper bounds for the EDM of the 
electron, of the neutron, and of the mercury $^{199}$Hg atom 
are~\cite{eedm,nedm,atomic}  
\begin{equation}
|d_e|< 4.23 \times 10^{-27}\;{\rm e}\:{\rm cm},\;\; 
|d_n|< 6.5 \times 10^{-26}\;{\rm e}\:{\rm cm}, \;\; 
C_{\rm Hg} < 3.0 \times 10^{-26}\;{\rm cm} \;,
\label{eq:EDM}
\end{equation}
where $C_{\rm Hg}$ is defined as in Ref.\cite{hg199}.
Large phases can be accommodated in several scenarios 
such as models with heavy sfermions~\cite{pran}, models 
with the cancellation mechanism~\cite{cancel},  
models with phases only in the third 
generation~\cite{chang}, or models with a 
non-trivial soft flavor structure~\cite{Abel}. 
Here, we use the cancellation 
mechanism~\cite{cancel}\footnote{For a more complete list 
of references and for a discussion of the effects of CP phases 
on low energy processes see Ref.\cite{cpreview}.}
which becomes possible if 
the SUGRA parameter space is extended to allow for 
different gaugino phases. The model we consider is thus 
described by the following parameters
\begin{equation}\label{parameters}
 m_0, m_{1/2}, \tan\beta, |A_0|, \theta_\mu, \alpha_0, 
\xi_1, \xi_2, \xi_3, 
\end{equation}
\noindent 
where, $\xi_i$ is the phase of the gaugino mass $M_i, i=1,2,3$. 
The value of $|\mu|$ is determined by imposing 
electroweak symmetry breaking (EWSB).

In the analysis we use a top-down approach, and thus 
impose Yukawa unification at the GUT scale, $M_{\rm GUT}$. 
For $b-\tau$ unification we have two independent  
Yukawa couplings at the grand unification scale, i.e., one common
$h_{\rm uni}$ for the b and the $\tau$,  
and one for the top-quark.  We use these to fit the experimental value of the 
 $\tau$  and the top masses. 
 Unless another value is specified, 
we fix the top mass at 178 GeV, 
which is its current experimental central value~\cite{mtop}. 
The value of $\alpha_s$ is fixed to be 0.1185.
For the $\tau$ mass at the electroweak 
scale $M_Z$, we use 1.7463 GeV, which takes 
into account the Standard Model radiative correction.
Naturally, we also take into account the SUSY correction, 
as derived in \cite{Ibrahim:2003ca}, when calculating 
$m_\tau$. In the case of the full Yukawa unification 
we impose $h_b=h_t=h_\tau=h_{\rm uni}$ at $M_{\rm GUT}$. 
Therefore, the value of $\tan\beta$ is fixed, since 
the two parameters $h_{\rm uni}$ and $\tan\beta$ 
are varied so as to obtain agreement with experimental 
values of $m_\tau$ and $m_{\rm top}$. 
As the b-quark couples to the same Higgs doublet ($H_d$) 
as the $\tau$ lepton, its mass is fixed by $h_{\rm uni}$. 
Therefore, $m_b(M_Z)$ is a prediction of our model and we 
require its value to be within the 2$\sigma$ range, 
\begin{equation}
  2.69\; {\rm GeV}\; < m_b(M_Z) \; < 3.10 \; {\rm GeV} 
\end{equation}
 as described in \cite{GLP1}.
In addition to the above, the other important constraints of the 
analysis are the relic density and the \bbsg constraint 
(see section \ref{secbsg}).

The procedure for the calculation of the particle and 
sparticle masses is as follows; After choosing a given 
set of the parameters in Eq.(\ref{parameters}), we run the 
renormalization group equations (RGEs) down to 
the SUSY scale, defined as the average of the 
two stop masses. At the SUSY scale the scalar potential 
is minimized and $|\mu|$ is calculated along with 
the SUSY threshold corrections to, e.g., the b quark  
and the $\tau$ lepton masses and the couplings are corrected 
accordingly. Hereafter, the sparticles are 
decoupled and the SM RGEs are used to run down to $M_Z$. 
At the electro-weak scale we check if the gauge couplings, the 
Weinberg angle, the top quark and the $\tau$ lepton masses 
are in agreement with their experimental values. If not, 
the RGEs are run iteratively until convergence is achieved.
In the analysis we use the two-loop SUSY renormalization group 
equations~\cite{Martin:1993zk} except for the trilinear terms, the
gaugino and sfermion masses, which are calculated at the 
one-loop level. The SUSY renormalization group equation will 
also be influenced by the CP-phases. However, it is easy to 
see that neither the phase of the $\mu$-term nor the phases 
of the gaugino masses will run. But, the phases of the 
trilinear terms run, and in general there will be three 
different phases at the low energy scale namely, $\alpha_t$, $\alpha_b$ 
and $\alpha_\tau$.  $\alpha_t$ is important as it affects 
$\Delta m_b$ as well as \bbsg. However, its value is almost  
fixed by the gluino phase. As shown in Ref.\cite{Carena:1994bv}, 
the approximate relation $A_{\rm top} \propto -M_3$ holds at low energy.

The regions of the mSUGRA parameter space that allows for 
acceptable relic abundance can be classified as: 
(i) the $\chi-\tilde \tau$ coannihilation region, 
(ii) the resonance region, and (iii) the Hyperbolic Branch/Focus Point (HB/FP) 
region\cite{ccn}. In a previous work~\cite{Gomez:2004ek}, we 
pointed out the strong variation of $\Delta m_b$ with CP phases. 
In that work we focussed on the effects 
induced by the SUSY corrections on the spectrum and 
their consequences for the neutralino relic density. 
It was shown that the CP-phases have a very large impact 
on the value of the CP-odd Higgs mass $M_A$, which in turn affects
 the predicted dark matter abundance in 
the so-called resonance region. The analysis of
Ref.\cite{Gomez:2004ek} used a bottom-up approach
by fixing the value of $m_b(M_Z)$ to its central value. 
In this work we use a top-down approach and large effects of the
CP phases are not seen. In fact, the predicted  neutralino relic abundance, 
turns out almost independent 
of the phases in the resonance region. In the stau 
coannihilation region there is also very little dependence 
on the CP-phases, except for the trilinear phase. 
As we show below, the HB/FP region cannot give rise 
to Yukawa unification within our model. 
In the calculation of the relic density we take into account the
CP even-CP odd Higgs mixing. In the MSSM, after spontaneous breaking
of the electro-weak symmetry one has at the tree-level two CP even Higgs 
($h^0,H^0$) and one CP odd Higgs ($A$). In the presence of CP 
violating phases these mix, producing mass eigenstates 
($H_1^0, H_2^0, H_3^0$), which are no longer eigen-functions of 
CP~\cite{cphiggsmass}\footnote{For further details regarding the 
implications of these CP even-CP odd Higgs mixings on neutralino 
dark matter analysis see Ref.\cite{Gomez:2004ek}.}

The most important supersymmetric threshold correction 
is the one to the bottom mass. 
At the loop level the effective b quark coupling with
the Higgs is given by~\cite{carena2002} 
\begin{equation}
 -{\cal {L}}_{bbH^0}= (h_b+\delta h_b) \bar b_R b_L H_1^0 + 
\Delta h_b \bar b_R b_L H_2^{0*} + H.c.
\end{equation}
The correction to the b quark mass is then given 
directly in terms of $\Delta h_b$ and $\delta h_b$ by  
\begin{equation}
\Delta m_b= [Re(\frac{\Delta h_b}{h_b}) \tan\beta 
+Re(\frac{\delta h_b}{h_b}) ].
\end{equation}
We use the full analysis of $\Delta m_b$ derived 
in~\cite{Ibrahim:2003ca}. 
The largest contributions to $\Delta m_b$ are the 
gluino and the chargino exchange contributions. The gluino exchange 
contribution is proportional to $M_3 \mu$, and will therefore 
depend on the phase combination $\theta_\mu+\xi_3$. 
The chargino exchange contribution is usually smaller, 
except for very large values of $|A_t|$, since it is 
proportional to $A_t \mu$. Its dominant phase dependence 
is given  by $\theta_\mu + \alpha_t$, 
and it has the opposite sign of 
the gluino contribution in a large region of the parameter space. 
When evaluating $h_b$ at $M_{\rm SUSY}$, we take into 
account threshold corrections using the relation
\footnote{This relation resums the SUSY self-energy 
leading order logarithmic corrections~\cite{Carena:1999py}.}
\begin{equation}
  h^{\rm SM}_b = h^{\rm SUSY}_b (1+\Delta m_b) \;.
\end{equation}  
The SM Yukawa coupling is evolved down to the electroweak 
scale, and the bottom quark mass
\begin{equation}
  m_b(M_Z)= h_b^{\rm SM} \frac{v}{\sqrt{2}} \cos \beta \;, 
\end{equation}
is calculated and compared with experiment. Similar 
expressions hold for the $\tau$ lepton with b replaced by $\tau$. 
For the top quark at the Z scale one has
\beqn
m_t(M_Z)=\frac{v}{\sqrt 2} \sin\beta h_t^{\rm SUSY} (1+\Delta m_t)
\eeqn
where
\begin{equation}
\Delta m_t= [Re(\frac{\Delta h_t}{h_t}) \cot\beta 
+Re(\frac{\delta h_t}{h_t}) ].
\end{equation}
A full analysis of $\Delta m_t$ is given in Ref.~\cite{Ibrahim:2003ca}.
However, in the region of interest which corresponds to large $\tan\beta$ 
the correction to the top quark Yukawa is essentially negligible.

\section{BR($b\rightarrow s \gamma$) with CP--phases}\label{secbsg}
The present average for the BR($b\rightarrow s \gamma$) derived  
from the available experimental data~\cite{bsgdata} is found to be,   
\begin{equation}
\rm{BR}(b\rightarrow s \gamma)=(3.54^{+0.30}_{-0.28})\times 10^{-4} \;,
\label{bsg1}
\end{equation}
by the {\it Heavy Flavor Averaging Group}~\cite{hfag}. 
The error includes an uncertainty due to the decay spectrum as 
well as the statistical error.
The theoretical SM prediction is~\cite{Gambino:2001ew,Buras:2002tp},
\begin{equation}
\rm{BR}(b\rightarrow s \gamma)=(3.70 \pm 0.30 )\times 10^{-4} \;.
\label{bsg2}
\end{equation}
The above result uses the $\overline{MS}$ running charm mass instead 
of the pole mass. It was claimed in Ref.~\cite{Gambino:2001ew} that 
this consideration reduces the NNLO uncertainty in the SM 
calculation. However, other analyses~\cite{Hurth,Neubert:2004dd} 
question the theoretical precision of Eq.(\ref{bsg2}) predicting a 
lower central value for the SM result. In any case, the result of 
Eq.(\ref{bsg2}) appears to be a good benchmark value for the SM 
prediction to work with.

The dominant SUSY contributions from the charged Higgs exchange 
include the $\tan\beta$ enhanced NLO corrections, which contribute 
to the Wilson coefficients $C_7$ and $C_8$ 
(these are coefficients of the operators 
$O_7=\frac{e}{16\pi^2} m_b (\bar s_L \sigma_{\mu\nu}b_R)F_{\mu\nu}$ and 
$O_8=\frac{g_s}{16\pi^2} m_b (\bar s_L \sigma_{\mu\nu} T^a b_R)G_{\mu\nu}^a$).
These contributions can be codified in  
$\epsilon_b^{'}(t)$, $\epsilon_t^{'}(b)$ and $\epsilon_{bb}$ which 
enter in the Lagrangian for effective interaction involving the 
charged Goldstone boson and the charged Higgs boson as follows
\beqn
{\cal {L}} & = & \frac{g}{\sqrt 2 M_W} G^+ \{ \sum_d m_t V_{td} \bar t_R d_L 
-\sum_u m_b V_{ub} \frac{1+\epsilon_b'(u)\tan\beta}{1+\epsilon_{bb}^*\tan\beta}
\bar u_L b_R\} \\
&+&\frac{g}{\sqrt 2 M_W} H^+ \{ \sum_d m_t V_{td} \bar t_R d_L 
\frac{1+\epsilon_t'(d)\tan\beta}{\tan\beta}
+ \sum_u m_b V_{ub}\bar u_Lb_R 
\frac{\tan\beta}{1+\epsilon_{bb}^*\tan\beta} \} + H.c. \nonumber
\eeqn
where $V_{ij}$ is the CKM mixing matrix.
Evaluation of $\epsilon_b^{'}(t)$, $\epsilon_t^{'}(b)$, and $\epsilon_{bb}$ 
exist in the literature~\cite{Degrassi:2000qf,Carena:2000uj},
but there is some ambiguity concerning the signs of some of the 
terms among the above groups.
To resolve this we carry out an independent analysis of these 
quantities for the same loop diagrams as in the previous works, 
including also their dependence on CP phases, which was
taken into account only in one analysis previously. 
Our analysis is derived using the work
of Ref.\cite{Ibrahim:2003tq}. We find 
\beqn
\epsilon_b^{'}(t) & = & 
-\sum_{i=1}^2\sum_{j=1}^2 \frac{2\alpha_s}{3\pi} e^{i\xi_3}D_{b2j}^*D_{t1i} 
[\frac{m_t}{m_b} \cot\beta A_t D_{b1j}D_{t2i}^*+ \mu D_{b2j}D_{t1i}^* 
+ m_t \cot\beta D_{b2j}D_{t2i}^* 
\nonumber\\
&& +\frac{m_t^2}{m_b} \cot\beta D_{b1j}D_{t1i}^* 
-\frac{m_W^2}{m_b} \sin\beta \cos\beta D_{b1j}D_{t1i}^* ]
\frac{1}{|m_{\tilde g}|} H(\frac{m_{\tilde t_i^2}}{|m_{\tilde g}|^2}, 
\frac{m_{\tilde b_j^2}}{|m_{\tilde g}|^2})
\nonumber \\
&& +2 \sum_{k=1}^4\sum_{i=1}^2\sum_{j=1}^2 
[\frac{m_t}{m_b} \cot\beta A_t D_{b1j}D_{t2i}^* 
+\mu D_{b2j}D_{t1i}^* + m_t \cot\beta D_{b2j}D_{t2i}^* 
\nonumber\\
&& +\frac{m_t^2}{m_b} \cot\beta D_{b1j}D_{t1i}^* 
-\frac{m_W^2}{m_b} \sin\beta \cos\beta D_{b1j}D_{t1i}^* ] 
\\
&& \times (\alpha_{bk}^* D_{b1j}^* - \gamma_{bk}^*D_{b2j}^*)
(\beta_{tk}D_{t1i}+\alpha_{tk}^* D_{t2i}) \frac{1}{16\pi^2} 
\frac{1}{m_{\chi_k^0}} H(\frac{m_{\tilde t_i^2}}{m_{ \chi_k^0}^2}, 
\frac{m_{\tilde b_j^2}}{m_{\chi_k^0}^2}) \nonumber
\label{bt}
\eeqn
In the above  $D_q$ is the matrix that diagonalizes the 
squark mass$^2$ matrix $M_{\tilde q}^2$, i.e., 
\beqn
D_q^{\dagger} M_{\tilde q}^2D_q ={\rm diag} 
(M_{\tilde q_1}^2, M_{\tilde q_2}^2)
\eeqn
and $H(a,b)$ is defined by 
\beqn
H(a,b)=\frac{a}{(1-a)(a-b)}\ln a + \frac{b}{(1-b)(b-a)}\ln b 
\eeqn
where $\alpha_{bk}, \beta_{bk},\gamma_{bk}$ for the b quark and 
the corresponding
coefficients for the t quark are as defined 
in Ref.\cite{Ibrahim:2003tq}.
\noindent
Similarly for $\epsilon^{'}_t(s)$ we find 
\beqn
\epsilon^{'}_t(s) &=& 
\sum_{i=1}^2\sum_{j=1}^2 \frac{2\alpha_s}{3\pi} e^{-i\xi_3}D_{s1i}^*D_{t2j} 
[\frac{m_s}{m_t} \tan\beta A_s^* D_{s2i}D_{t1j}^* + 
\mu^* D_{s1i}D_{t2j}^* + m_s \tan\beta D_{s2i}D_{t2j}^* 
\nonumber\\
&& +\frac{m_s^2}{m_t} \tan\beta D_{s1i}D_{t1j}^* 
-\frac{m_W^2}{m_t} \sin\beta \cos\beta D_{s1i}D_{t1j}^* ]
\frac{1}{|m_{\tilde g}|} H(\frac{m_{\tilde s_i}^2}{|m_{\tilde g}|^2}, 
\frac{m_{\tilde t_j}^2}{|m_{\tilde g}|^2})
\nonumber\\
&& -2 \sum_{k=1}^4\sum_{i=1}^2\sum_{j=1}^2 
[\frac{m_s}{m_t} \tan\beta A_s^* D_{s2i}D_{t1j}^*+ 
\mu^* D_{s1i}D_{t2j}^* + m_s \tan\beta D_{s2i}D_{t2j}^* 
\nonumber\\
&& +\frac{m_s^2}{m_t} \tan\beta D_{s1i}D_{t1j}^* 
-\frac{m_W^2}{m_t} \sin\beta \cos\beta D_{s1i}D_{t1j}^* ]
\\
&& \times (\beta_{sk}^* D_{s1i}^*+\alpha_{sk} D_{s2i}^*)
(\alpha_{tk} D_{t1j} - \gamma_{tk} D_{t2j} )
\frac{1}{16\pi^2} 
\frac{1}{m_{\chi_k^0}} H(\frac{m_{\tilde s_i}^2}{m_{ \chi_k^0}^2}, 
\frac{m_{\tilde t_j}^2}{m_{\chi_k^0}^2}) \nonumber
\label{epsilonts}
\eeqn
Finally, our analysis of $\epsilon_{bb}$ gives
\beqn
\epsilon_{bb}&=&-\sum_{i=1}^2\sum_{j=1}^2 \frac{2\alpha_s}{3\pi} e^{-i\xi_3}
D_{b1i}^*D_{b2j} 
[\frac{M_Zm_W}{m_b} \frac{\cos\beta}{\cos\theta_W}
\{(-\frac{1}{2} +\frac{1}{3}\sin^2\theta_W)D_{b1i}D_{b1j}^*
\nonumber\\
&& -\frac{1}{3}\sin^2\theta_W D_{b2i}D_{b2j}^*\}\sin\beta
+\mu^* D_{b1i}D_{b2j}^* ]
\frac{1}{|m_{\tilde g}|} H(\frac{m_{\tilde b_i}^2}{|m_{\tilde g}|^2}, 
\frac{m_{\tilde b_j}^2}{|m_{\tilde g}|^2})
\nonumber\\
&& -\sum_{i=1}^2\sum_{j=1}^2 \sum_{k=1}^2 
g^2 [\frac{M_Zm_W}{m_b} \frac{\cos\beta}{\cos\theta_W} 
\{(\frac{1}{2} -\frac{2}{3}\sin^2\theta_W) D_{t1i}  D_{t1j}^*
+ \frac{2}{3}\sin^2\theta_W  D_{t2i}D_{t2j}^*\}\sin\beta
\nonumber\\
&& -\frac{m_t^2}{m_b} \cot\beta 
\{D_{t1i}D_{t1j}^* + D_{t2i} D_{t2j}^{*}\} 
- \frac{m_t}{m_b} \cot\beta A_t^* D_{t2i}D_{t1j}^* ]
\\
&& \times (V_{k1}^*D_{t1i}^*-K_t V_{k2}^* D_{t2i}^* ) (K_b U_{k2}^* D_{t1j}) 
\frac{1}{16\pi^2} \frac{1}{|m_{\tilde \chi_k^+}|} 
H(\frac{m_{\tilde t_i}^2}{|m_{\tilde\chi_k^+}|^2}, 
\frac{m_{\tilde t_j}^2}{|m_{\tilde \chi_k^+}|^2}) 
\label{bb}
\eeqn
The form factor $H(a,b)$ in the above equation can have $a=b$ 
and in this case it reads
\beqn
H(a,a)=\frac{1}{(a-1)^2}[1-a+\ln a]
\eeqn
Before proceeding further we give a brief comparison of these 
results with the results of the previous works. The analysis 
of $\epsilon^{'}_b(t)$ may be compared to $\epsilon_{tb}$ of 
Ref.\cite{do} in the limit of large $\tan\beta$ and small squark 
mixings. In this case the limit of the first two lines in Eq.(\ref{bt}) 
agrees with the result of Ref.\cite{do}. However, the limit of the last
three lines of Eq.(\ref{bt}) have an opposite sign to that of 
Ref.\cite{do}. Here our analysis is in
agreement with the result of Ref.\cite{micro}. 

Next we give a computation of $\epsilon^{'}_t(s)$. 
Approximating Eq.(\ref{epsilonts}) we find 
\beqn
\epsilon^{'}_t(s) 
=\sum_{i=1}^2\sum_{j=1}^2 \frac{2\alpha_s}{3\pi} e^{-i\xi_3}
\mu^* |D_{s1i}|^2 |D_{t2j}|^2 
\frac{1}{|m_{\tilde g}|} H(\frac{m_{\tilde s_i^2}}{|m_{\tilde g}|^2}, 
\frac{m_{\tilde t_j^2}}{|m_{\tilde g}|^2})
\nonumber\\
-\frac{h_s^2}{16\pi^2} \frac{A_s^*}{m_{\chi_k^0}} X_{3k}X_{4k}  
|D_{s2i}|^2 |D_{t1j}|^2   H(\frac{m_{\tilde s_i^2}}{|m_{\tilde \chi_k^+}|^2}, 
\frac{m_{\tilde t_j}^2}{|m_{\tilde \chi_k^+}|^2})
\label{ts}
\eeqn
The analysis of Ref.\cite{do} computed only the first line of 
Eq.(\ref{ts}) and for this part we agree with their work when 
we take the large $\tan\beta$ limit and the limit of small 
mixing angles of our result. The work Ref.\cite{Degrassi:2000qf} 
gives results  corresponding to Eq.(\ref{ts}). However, here we find
that we have a disagreement with the sign of the second part of their Eq.(16).

Our analysis of $\epsilon_{bb}$ given by our Eq.(\ref{bb}) agrees 
with the analysis of Ref.\cite{do} in the
limit of large $\tan\beta$ and in the limit of small squark 
mixings and here there is a general agreement 
(taking account of typo corrections) among various groups in the 
limit of no CP phases. Our 
analysis like that  of Ref.\cite{do} takes into account the full
dependence on CP phases. 
In the numerical analysis to be presented below, we have used 
the code provided by micrOMEGAs \cite{micro} in the CP conserving
case. This code agrees with the codes used by other 
groups~\cite{Ellis:2004tc}. In the CP violating case we have combined 
the codes of Refs.\cite{micro, ms} with our own 
codes of the SUSY contributions with CP phases.

For the uncertainty in BR($b\rightarrow s \gamma$) we use a linear 
combination of the errors on Eqs.~(\ref{bsg1}) and (\ref{bsg2}). 
At the 2-$\sigma$ level one has
\begin{equation}
2.3\times 10^{-4}<\rm{BR}(b\rightarrow s \gamma)< 4.7\times 10^{-4} \;.
\label{bsg3}
\end{equation}  
The numerical analysis given below is controlled essentially by 
the upper bound in Eq.(\ref{bsg3}). 
 In order to obtain the correct value of $m_b(m_Z)$ one 
needs the phase combination $\theta_\mu + \xi_3$ to be close to 
$\pi$\footnote{The phase combination is drawn to smaller values 
for large $\tan\beta$ and for the full Yukawa unification 
it ends up close to $\pi/2$.}. In this case the chargino contribution to  
the \bbsg is positive and therefore the lower bound is not 
reached for the values of the SUSY parameters in our study.
We turn now to the details of the numerical analysis.

\section{WMAP Dark Matter Constraints and 
$b-\tau$ Yukawa unification}
\begin{figure}
\centering
\includegraphics[width=9cm,height=7cm]{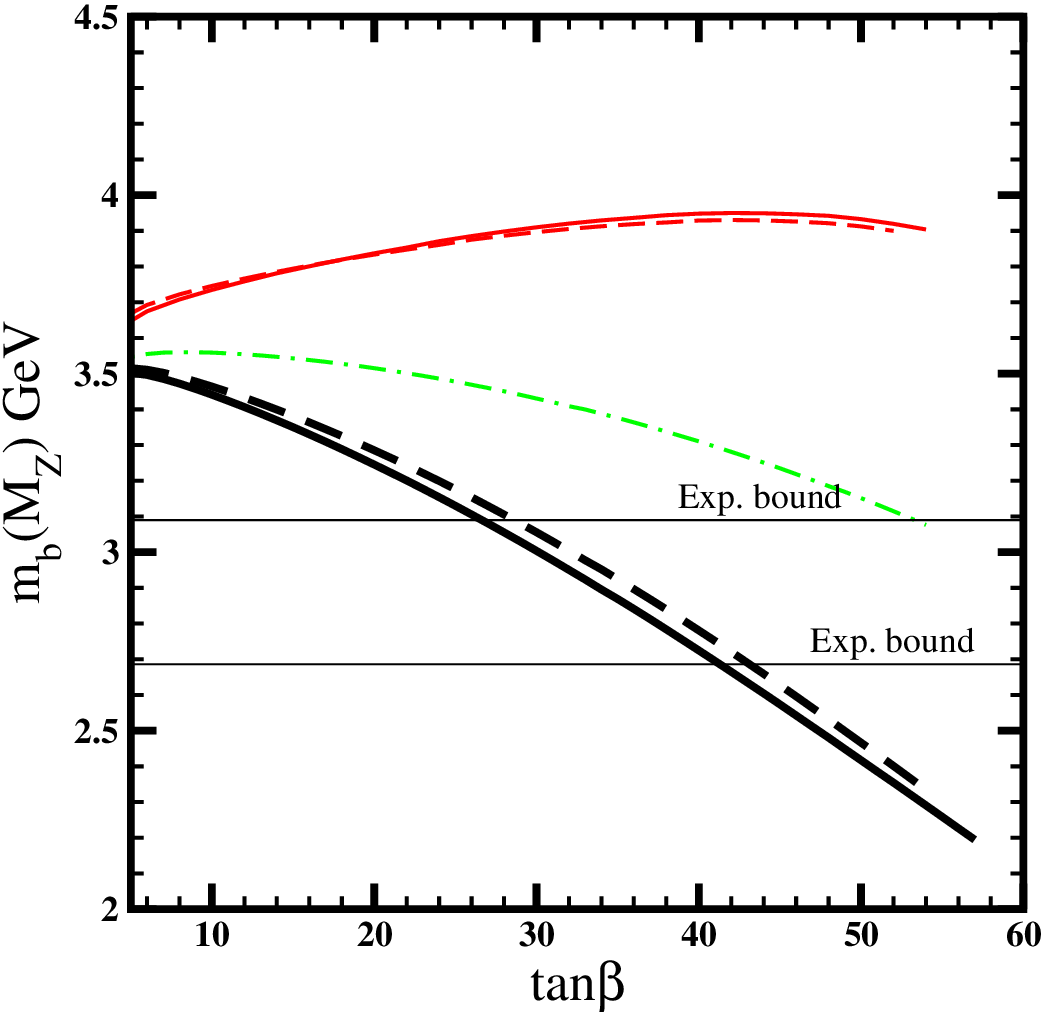}
\caption{ The value of $m_b(M_Z)$ versus $\tan\beta$ 
assuming $h_b=h_\tau$ at the GUT scale with $m_{1/2}=800 \; \rm{GeV}, 
A_0=0 \; \rm{GeV}$, $m_0=300 \; \rm{GeV}$ (solid lines),
$m_0=600 \; \rm{GeV}$ (dashed lines). The thick (thin) lines 
have $\mu<0$ ($\mu>0$). The dot-dashed line is plotted with $\Delta m_b=0$. 
}
\label{mbtb_ab}
\end{figure}

\begin{figure}
\centering
\includegraphics[width=9cm,height=7cm]{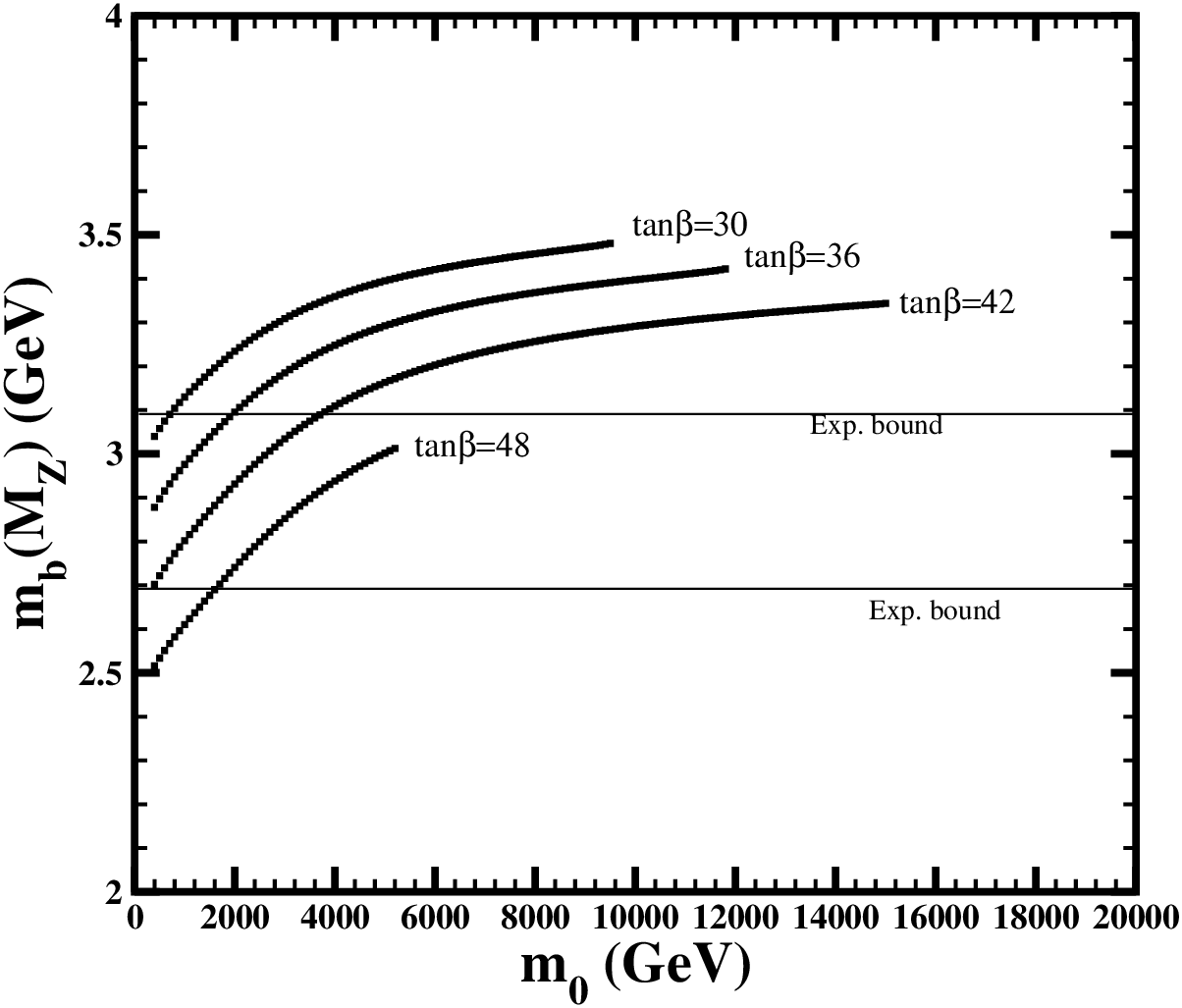}
\caption{The value of $m_b(M_Z)$ using the constraint of $b-\tau$ 
Yukawa unification in mSUGRA. Furthermore, $m_t=176$~GeV, 
$\mu<0$ ($\theta_\mu=\pi$), $m_{1/2}=800$~GeV, $A_0=0$ and the 
value of $\tan\beta$ has been varied as indicated 
on the curves. Lines ends at values of $m_0$ where 
EWSB is no longer satisfied,  except for $\tan\beta=48$, 
where the CP odd Higgs mass is too small.
}
\label{focusmb}
\end{figure}

It is useful to first summarize our results in the mSUGRA case, 
where all CP phases are either zero or $\pi$.  
The relation $h_b=h_\tau$ can be satisfied for a wide 
range of soft masses in the MSSM with real universal soft 
terms. To discuss the dependence of $m_b(M_Z)$ on $\tan\beta$ 
we consider two representative set of soft parameters: 
(i) $m_{\frac{1}{2}}=800$ GeV, $A_0=0$, $m_0=300$ GeV, and 
(ii) $m_{\frac{1}{2}}=800$ GeV, $A_0=0$, $m_0=600$ GeV. 
In Fig.~(\ref{mbtb_ab}) we study the $\mu>0$ and $\mu<0$ cases  
for each set. The lines corresponding to case (i) are 
interrupted when $m_{\tilde{\tau}}<m_\chi$, while for case (ii)  
large values of $\tan\beta$ are incompatible with EWSB. 
We include a reference line ignoring the SUSY threshold corrections 
(i.e., $\Delta m_b=0$). Fig.~(\ref{mbtb_ab}) exhibits the well known 
phenomenon, that $\Delta m_b$ is positive for $\mu$ positive and
therefore the theoretical prediction for the b quark pole mass is 
too high, lying outside the experimental range. Thus b-$\tau$ 
unification does not occur in this case. When $\mu<0$, on the other
hand, $\Delta m_b$ is negative and the theoretical prediction 
for the b quark mass can lie within the experimental range 
for values of $\tan\beta$ between roughly 25 and 45. 
A similar analysis of $m_b(M_Z)$ but as a function of $m_0$ is given in 
Fig.(\ref{focusmb}). Here we consider only the $\mu<0$ case and find
that the theoretical prediction of $m_b(M_Z)$ can lie within the
corridor allowed by experiment for a range of $m_0$ values. However, one
finds that for very high values of $m_0$, i.e., 
for values above 5 TeV and beyond, a region which 
includes 
the Hyperbolic Branch (HB)/Focus Point (FP) region, Yukawa 
unification is not achieved with universal soft parameters.

\begin{figure}[h]
\centering
\includegraphics[width=12cm,height=8cm]{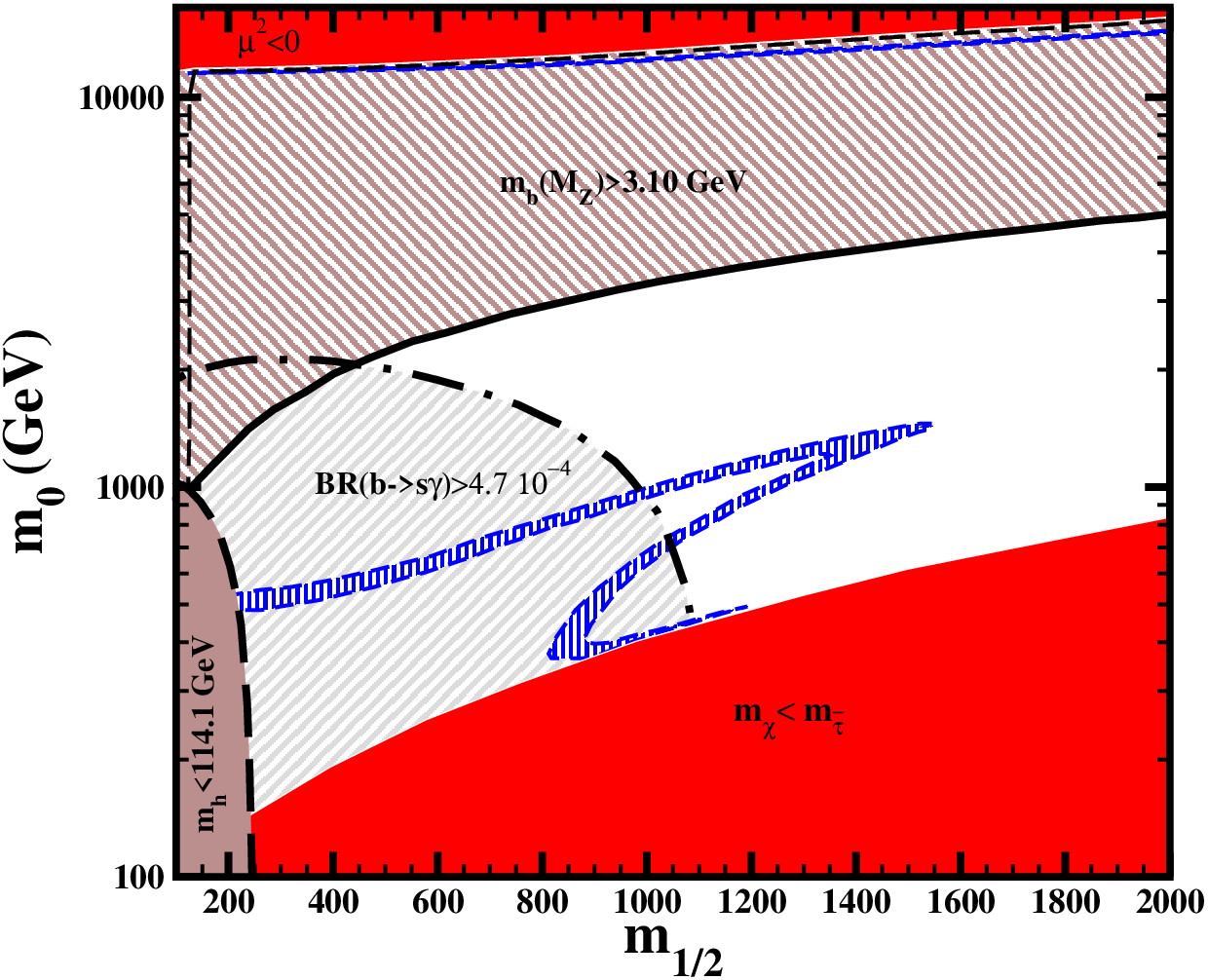}
\caption{Analysis 
of neutralino relic density, \bbsg and of $m_b(M_Z)$ 
including the HB/FP region for $\tan\beta=40$, 
$A_0=0$ and $m_t=176$~GeV. Areas contoured by the dashed line 
has a neutralino relic density is inside the WMAP bounds. 
The area above the solid line predicts $m_b(M_Z)>3.10$~GeV 
while the area inside the dashed (dot--dashed) line is 
excluded by the lower bound on $m_h$ (the upper bound on 
$BR(b\rightarrow \gamma)$). On the lower dark shaded area 
$m_\chi > m_{\tilde{\tau}}$ while on the upper EWSB is not 
achieved. The thiner dashed line indicates $m_{\chi^+}=103$~GeV.
}
\label{alldm}
\end{figure}

We extend the analysis now to include the relic density constraints. 
Fig.(\ref{alldm}) shows an area plot in the $m_0-\mhalf$ plane 
and all the three interesting dark matter regions in mSUGRA 
can be seen. In Fig.(\ref{alldm}) the coannihilation area 
and the resonance area overlap. The HB/FP area, which is the 
region adjacent to the area with no EWSB ($\mu^2<0$), is 
incompatible with any kind of Yukawa unification within 
the framework of universality of soft parameters at the GUT 
scale as already seen in the analysis of Fig.(\ref{focusmb}). 

The HB/FP region moves to lower values of $m_0$ as  
$tan\beta$ and $m_t$ decrease and this variation, especially 
with $m_t$, can be very large 
 (we present our analysis in Figs.(\ref{focusmb},\ref{alldm})
 using $m_t=176$~GeV in order that the HB/FP area appear 
below 20 TeV).
With $m_t$ at its lower bound the HB/FP region appears at 
values of $m_0$ of 5-6 TeV, but here $m_b$ is already too high. 
With large values of $tan\beta$, $m_A^2$ becomes 
negative before $\mu$ becomes small and therefore 
there is no inversion of the gaugino/Higgsino components  
in the composition of $\chi^0$. Correspondingly, 
the HB/FB is not reached. This is the case for the line 
of $\tan\beta=48$ in Fig.\ref{focusmb}.  
Therefore, overlapping of the HB/FP region and the 
allowed $m_b$ area is not possible.
Moreover, phases cannot improve the situation as 
$\Delta m_b$ is very small in the HB/FP (below 5\%), and 
thus cannot lower the value of $m_b$ sufficiently.

\begin{figure}
\centering
\includegraphics[width=7cm,height=7cm]{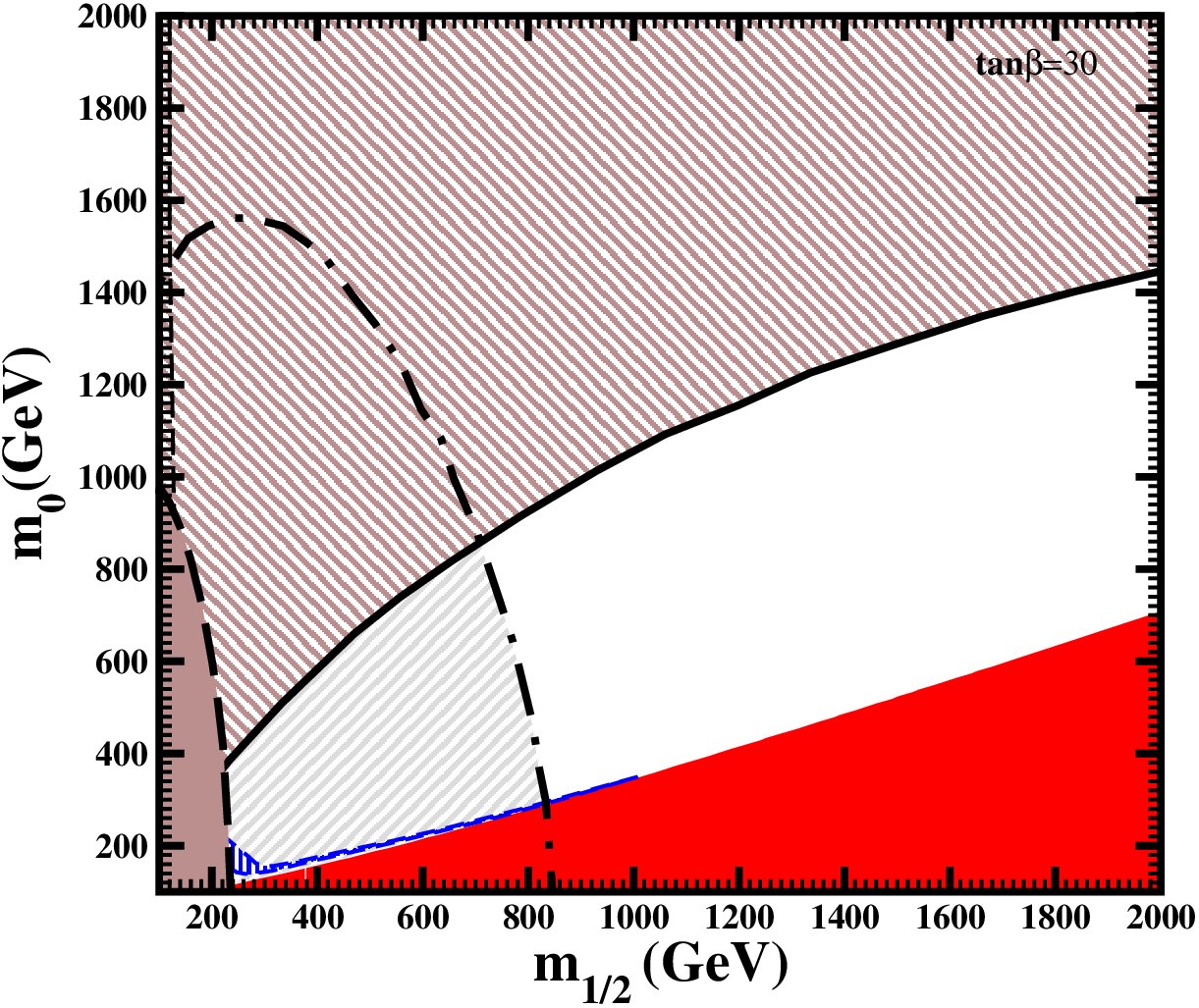}
\includegraphics[width=7cm,height=7cm]{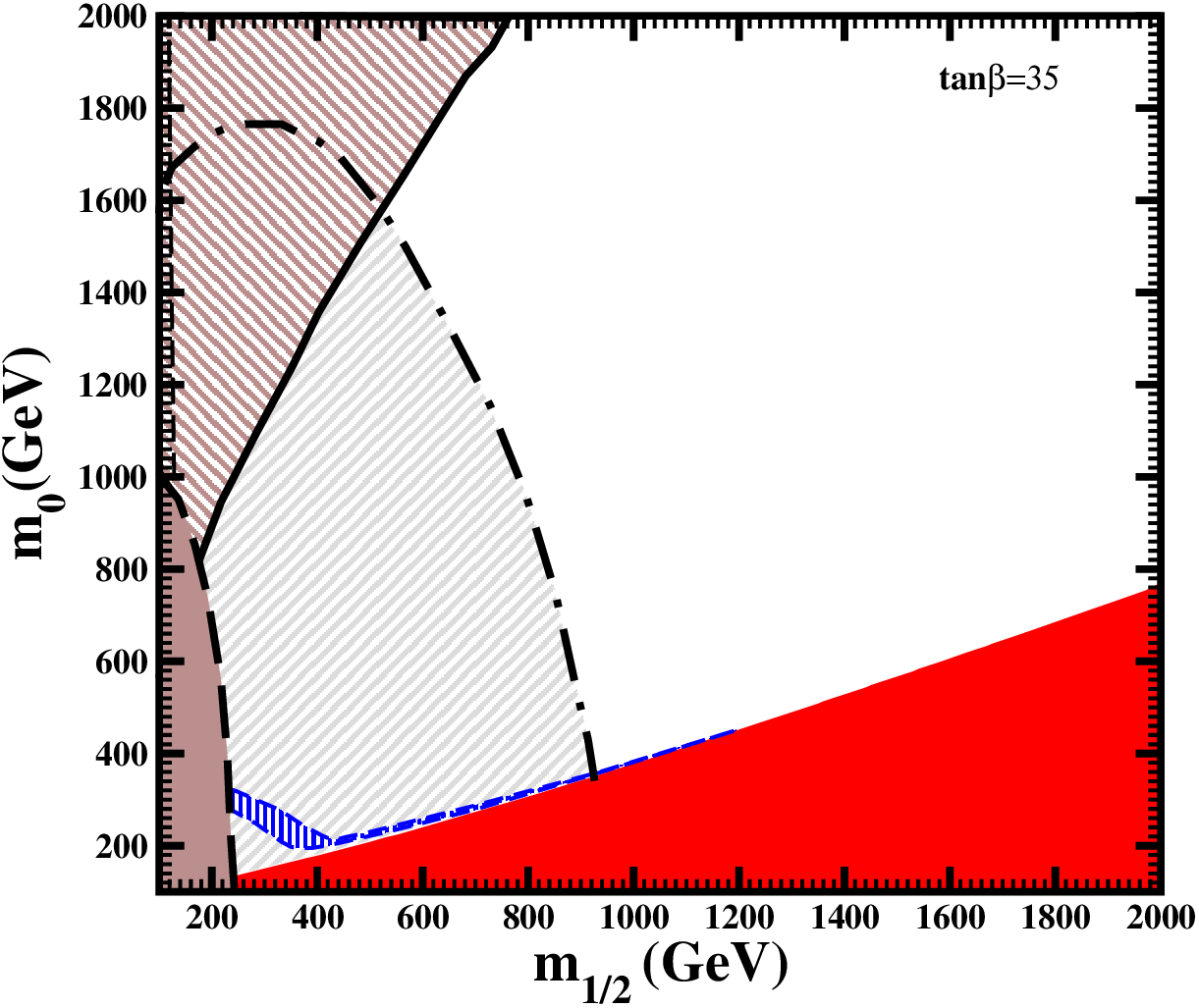}
\includegraphics[width=7cm,height=7cm]{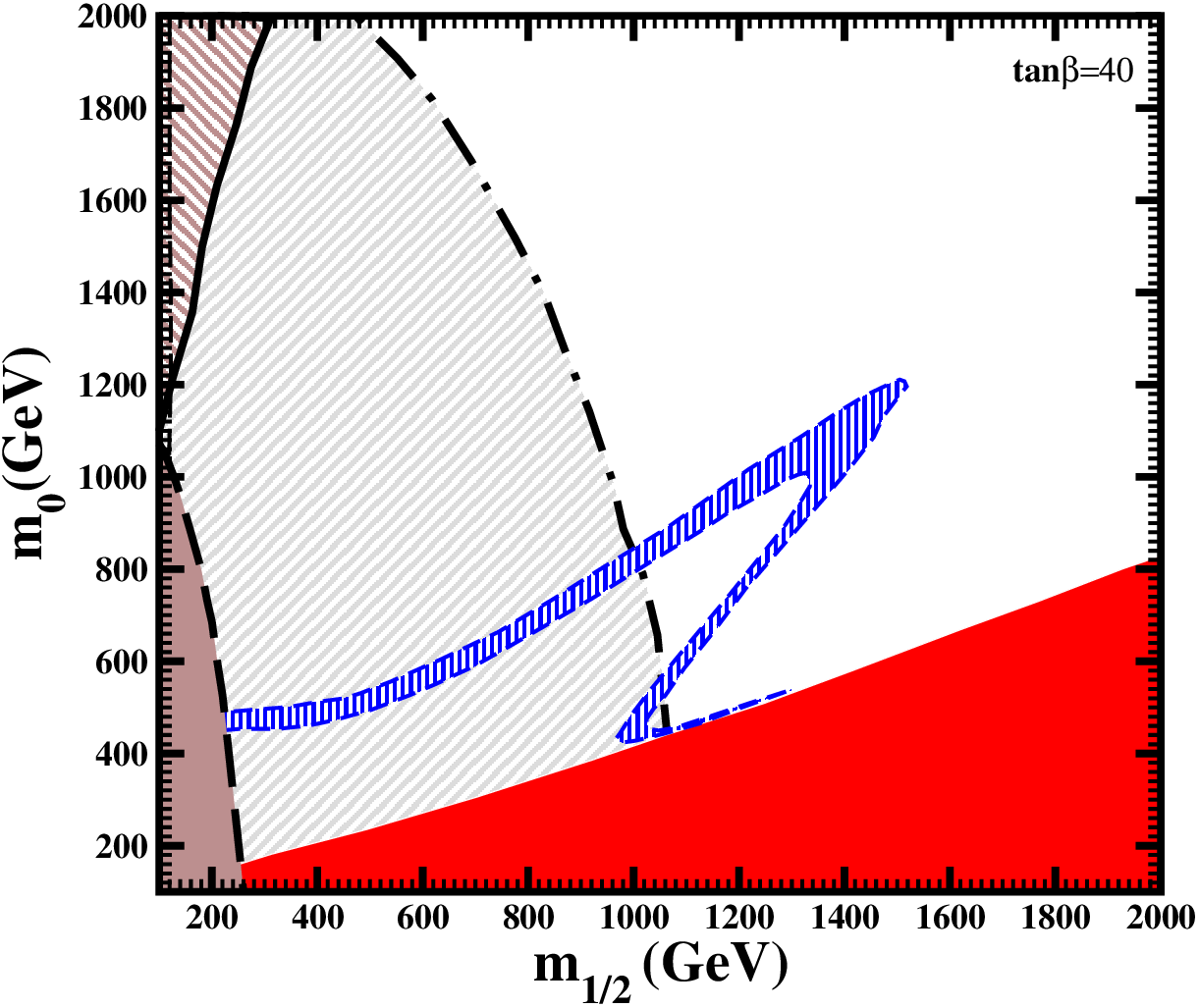}
\includegraphics[width=7cm,height=7cm]{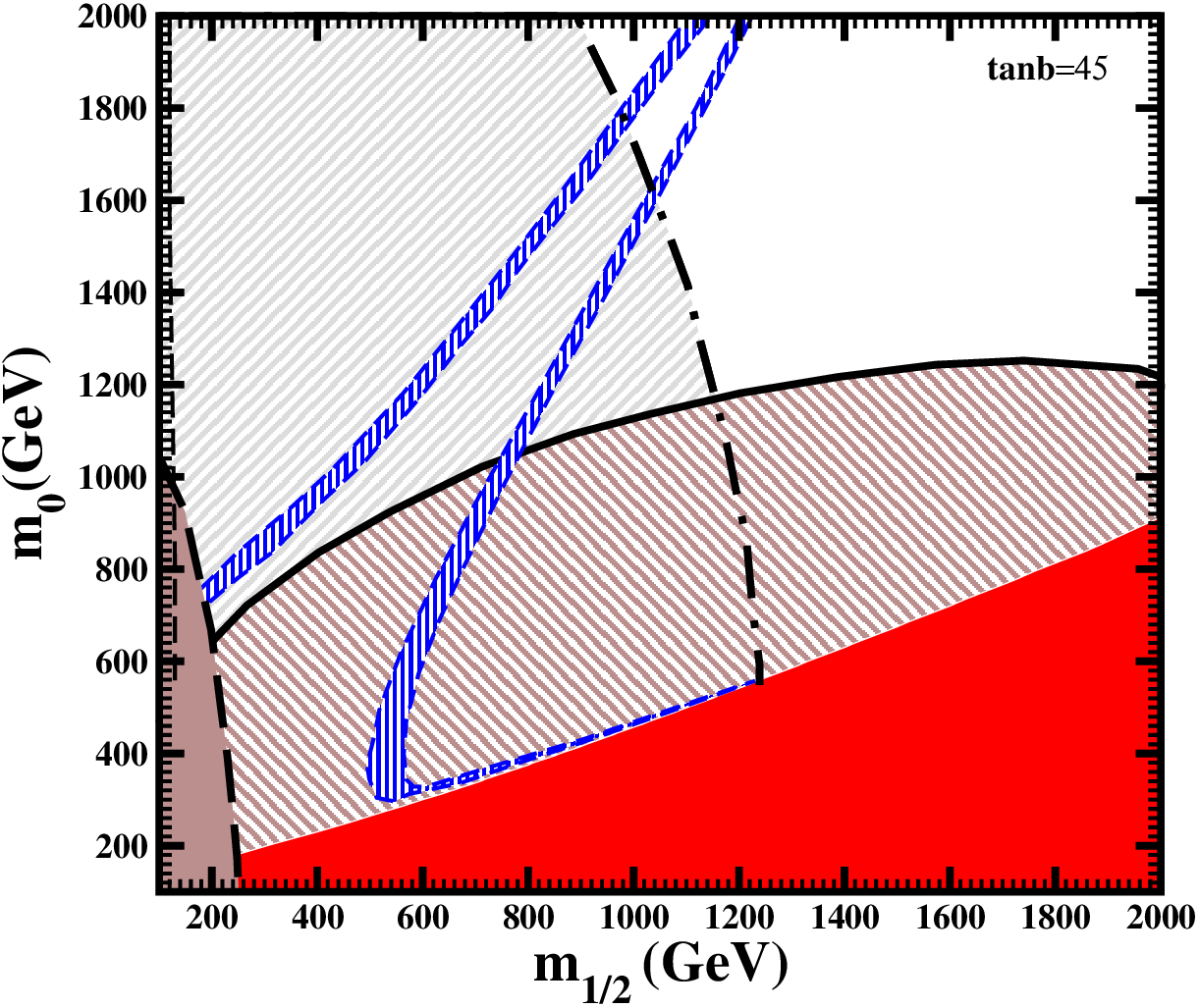}
\caption{Analysis of the neutralino
relic density with the $b-\tau$ Yukawa unification constraints 
in the $m_0-M_{1/2}$ plane when the soft terms are universal and real 
with $\mu<0$ ($\theta_\mu=\pi$), $A_0=0$, $m_t=178$~GeV,and 
$tan\beta=30, 35, 40, 45$.  
The lines and shaded areas are as described in Fig.\ref{alldm}. 
The area inside the solid line 
for the case $tan\beta=45$ predicts $m_b(M_Z)<2.69$~GeV.
}
\label{mom12_su5}
\end{figure}

In Fig.~(\ref{mom12_su5}) we further analyze the mSUGRA case 
with area plots in the $m_0-m_{\frac{1}{2}}$ plane for 
four values of $\tan\beta$: 30, 35, 40, 45.
Fig.~(\ref{mom12_su5}) shows that the relic abundance, 
$b-\tau$ unification, and $b\rightarrow s \gamma$ constraints 
can be simultaneously satisfied for a narrow range of parameters 
for values of $\tan\beta$ in the range 30--45.
The \bbsg constraint is a major restriction, since
       both SUSY and Higgs contributions to the branching ratio
       add to the one from the Standard Model, and thus one needs a relatively
       heavy spectrum such that the \bbsg prediction remains below the
        experimental upper bound.
The b-$\tau$ unification constraint and the WMAP
constraint further reduce the parameter space. Even so, 
one finds  that there exist regions of the parameter space  
for all the four cases in Fig.~(\ref{mom12_su5}) consistent 
with the WMAP data under the $b-\tau$ unification and 
$b\rightarrow s \gamma$ constraints.

\subsection{Effects of CP phases}
To determine the impact of phases on the above picture we 
choose two representative points from Fig.~\ref{mom12_su5}:
\begin{equation}
a)\;\; tan\beta=30, m_0= 290\; \rm{GeV}, 
\; M_{1/2}=800\; \rm{GeV}, \; A_0=0 \; \rm{GeV}  
\label{point30}
\end{equation}
\begin{equation}
b)\;\;  tan\beta=40, m_0= 710 \; \rm{GeV}, \; M_{1/2}=800\; \rm{GeV}, \; 
A_0=0 \;\rm{GeV}
\label{point40}
\end{equation}
These points are chosen because in the absence of phases the WMAP relic
density constraints are satisfied by different mechanisms 
for these two cases. Thus, for the point in Eq.(\ref{point30}) 
the WMAP constraint is satisfied due to $\chi-\tilde\tau$
coannihilations. In contrast, for the point in Eq.(\ref{point40}) 
the WMAP constraint is satisfied due to a resonance in the 
Higgs mediated annihilation of $\chi-\chi$.   
To determine the effect of phases we study the most 
relevant phases for the processes that we consider. 
 The phases $\xi_1$ and $\xi_2$ have little impact on 
$b \rightarrow s \gamma$ and $\Delta m_b$. For simplicity we 
set them to zero in this section.
We have already discussed the phase combinations that play an important role
in the analysis  of  $\Delta m_b$. For the analysis of $b \rightarrow s \gamma$ we find
that the same phase combinations, i.e, $Arg(\mu A_t)$ and $Arg(\mu M_3)$ are
the important ones.

\begin{figure}
\hspace*{-0.6in}
\centering
\includegraphics[width=9cm,height=7cm]{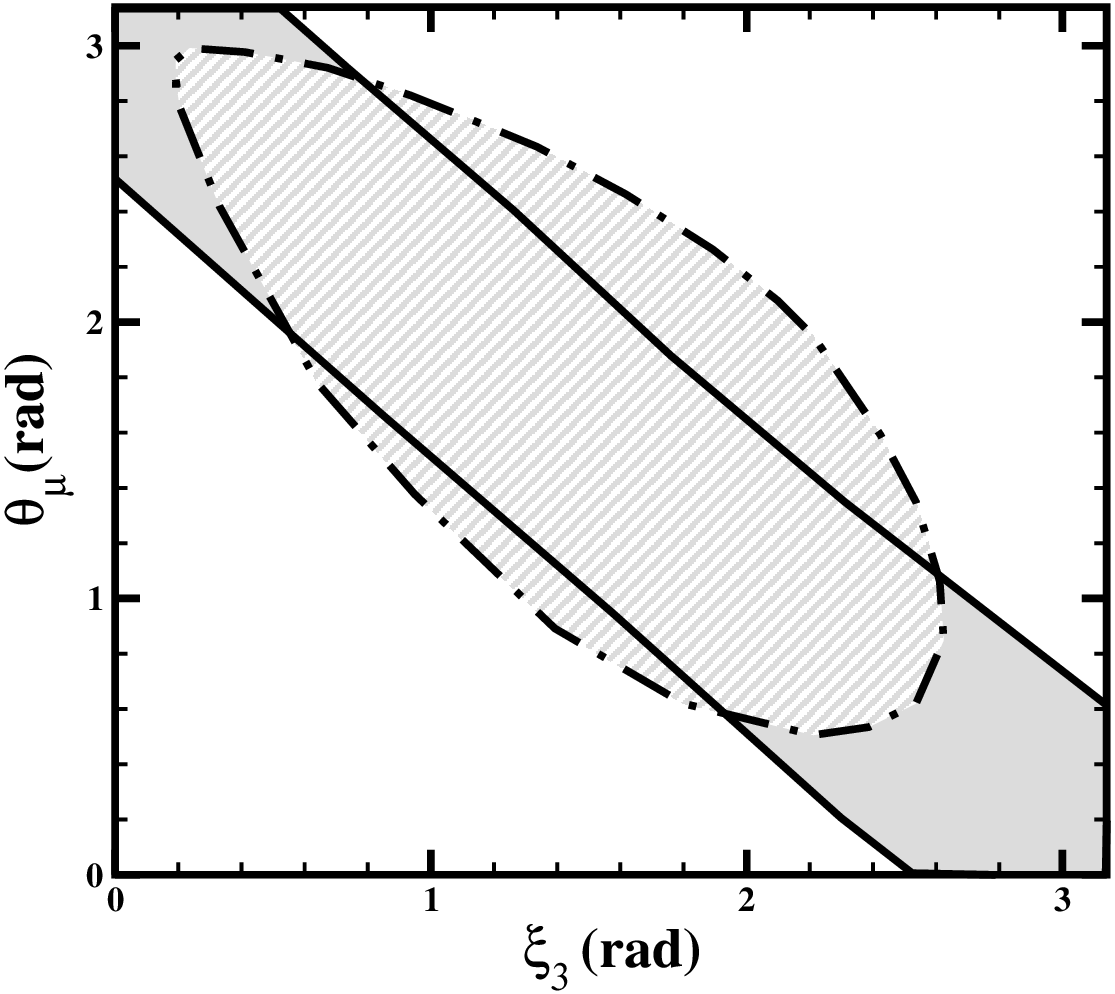}
\caption{ A plot in the $\xi_3-\theta_\mu$ plane for $\tan\beta=30$, 
$m_0=290$~GeV, $m_{1/2}=800$~GeV, $A_0=0$ and $\xi_1=\xi_2=0$. 
The area inside the solid lines predicts $m_b(M_Z)$ within 
the 2-$\sigma$ experimental range. The area inside the dot-dashed 
line is excluded by the BR($b\rightarrow s \gamma$) constraint.  
}
\label{xi3thetamu_30}
\end{figure}

\begin{figure}
\hspace*{-0.6in}
\centering
\includegraphics[width=9cm,height=7cm]{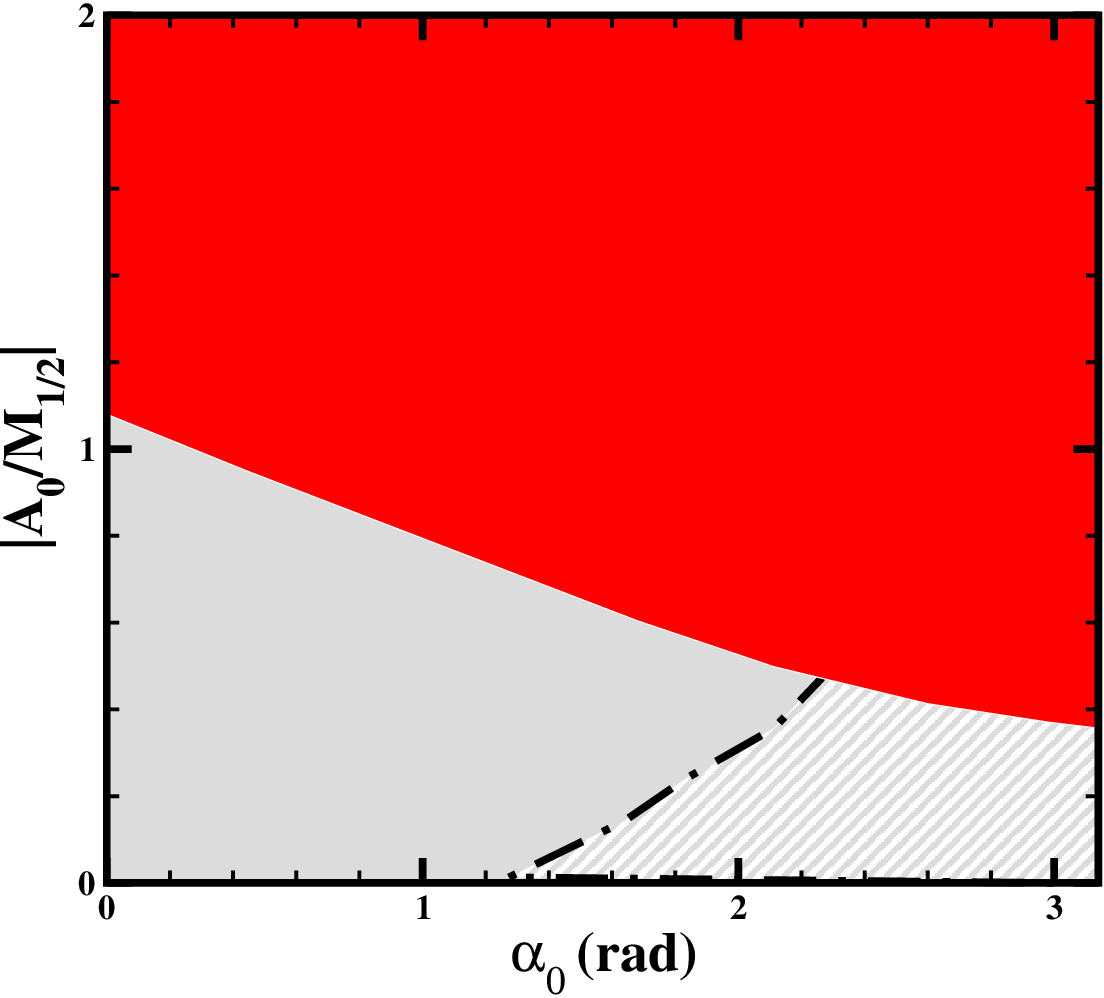}
\caption{Analysis of $b-\tau$ unification in the 
$|A_0/m_{1/2}|-\alpha_0$ plane with phases corresponding to the point 
in Eq.(\ref{point30}) with $\xi_3=0.3$~rad, $\theta_\mu=2.4$~rad and 
$\xi_1=\xi_2=0$. In the dark shaded area, $m_\chi > m_{\tilde{\tau}}$ 
while within the ruled area bounded by the dot-dashed line 
BR($b\rightarrow s \gamma$) exceeds its upper limit. 
The grey area predicts $m_b(M_Z)$ inside the experimental
bounds and fills the whole plane.}
\label{a0pha_3180}
\end{figure}

We now discuss the specifics of the point in Eq.~(\ref{point30}), 
which as already stated is in the $\chi-\tilde{\tau}$ 
coannihilation region.
In Fig.(\ref{xi3thetamu_30}) we analyze the \bbsg and the $b-\tau$ 
unification constraints in the $\theta_{\mu}-\xi_3$ plane, and 
as is seen the point satisfies the $b \rightarrow s \gamma$ as well 
as $m_b(M_Z)$ constraint in the mSUGRA case with a negative $\mu$.
The figure illustrates that the inclusion of phases 
changes the value of $m_b(M_Z)$ and \bbsg drastically. 
However, the two above mentioned constraints have a tendency 
to conflict with each other, even with the inclusion of CP phases. 
Nevertheless, we find that there exists a substantial overlap 
of the areas allowed by the bounds on $m_b(M_Z)$ and \bbsg.  
At the same time the predicted
$\Omega_{CDM} h^2$ remains inside the WMAP bounds because the phases do 
not affect significantly the ratio $m_\chi/m_{\tilde{\tau}}$ and hence 
the relic density prediction remains dominated by coannihilations. 
It is also  instructive to study the effects of $\alpha_0$. 
In Fig.~(\ref{a0pha_3180}) we analyze the dependence on 
$|A_0|$ and $\alpha_0$ for the point $\xi_3=0.3$ rad 
and $\theta_\mu=2.4$ rad of  Fig.(\ref{xi3thetamu_30}).  
In the analysis of Fig.~(\ref{a0pha_3180}) the ratio 
$ m_{\tilde{\tau}}/m_\chi$ does not exceed 1.08, and thus we 
remain in the coannihilation region allowing for the 
satisfaction of the relic density constraints. 
Furthermore, it is also possible to satisfy 
BR($b\rightarrow s \gamma$)  and $m_b$ bounds 
simultaneously for a wide range of $|A_0|$ and $\alpha_0$.

\begin{figure}
\centering
\includegraphics[width=9cm,height=7cm]{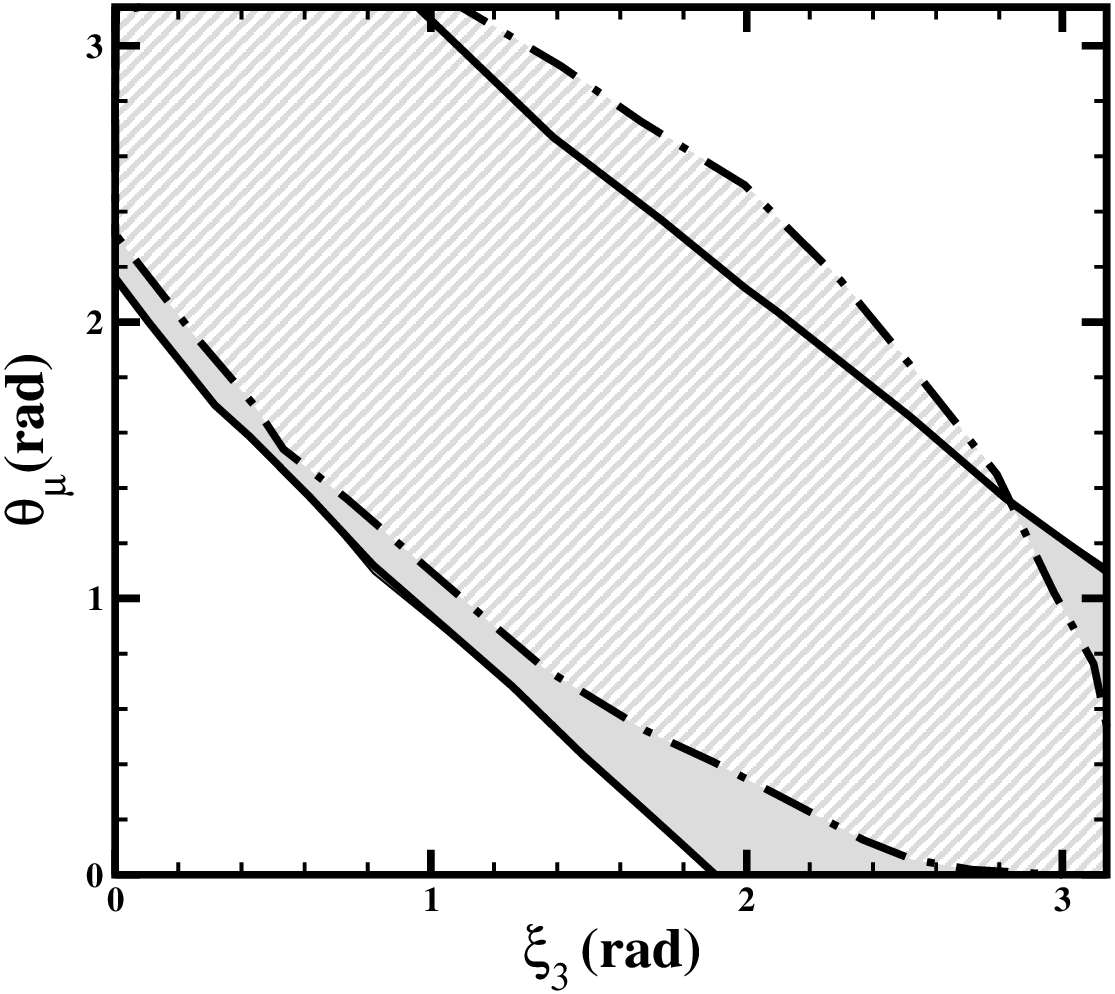}
\caption{ Same as Fig.\ref{xi3thetamu_30} 
except that $tan\beta=40$, $m_0=710$~GeV, 
$m_{1/2}=800$~GeV, $A_0=\xi_1=\xi_2=0$
}
\label{xi3thetamu_40}
\end{figure}

Next we analyze the implication of phases for the 
point in Eq.~(\ref{point40}). As already indicated
this point is within the resonance region in the mSUGRA case. 
However, the point produces a value for 
the BR($b\rightarrow s \gamma$) outside the experimental 
bounds as may be seen from Fig.~(\ref{mom12_su5}). 
The effect of varying $\xi_3$ and $\theta_\mu$ is analyzed 
in Fig.~(\ref{xi3thetamu_40}). 
Here one finds a substantial overlap of the areas allowed 
by the bounds on $m_b(M_Z)$ and BR($b\rightarrow s \gamma$) 
while the relic density prediction remains within the WMAP bounds.
As already stated, this analysis is substantially different from the one given 
in Ref.~\cite{Gomez:2004ek} at $\tan\beta=40$. There 
$m_b(M_Z)$ was fixed and the dependence of $\Delta m_b$ on the phases 
has a big effect on the resonant channels. For the present case, the 
bottom Yukawa has only a small fluctuation due to the unification condition 
at the GUT scale. Thus its effect on the Higgs mass 
parameters through the RGE's is not as large as the 
one found in the analysis of Ref.~\cite{Gomez:2004ek}. 
Thus in the analysis of Ref.~\cite{Gomez:2004ek} no 
unification condition was assumed for the Yukawa couplings, 
and the only requirement on them was to predict fixed values 
for the fermion masses. 
In the case of $m_b$, the effects induced by the phases via 
$\Delta m_b$ were compensated by variations on $h_b$ so as 
to obtain a fixed $m_b(M_Z)$. Since such adjustments of 
$h_b$ induced large changes on the Higgs parameters, 
the relic density was very sensitive to  the phases. 
In the present case, $h_b$ is approximately fixed by the 
condition $h_b=h_\tau$ at the GUT scale (where $h_\tau$ is 
determined by $\tan\beta$ and $m_\tau$).

\begin{figure}
\hspace*{-0.6in}
\centering
\includegraphics[width=9cm,height=7cm]{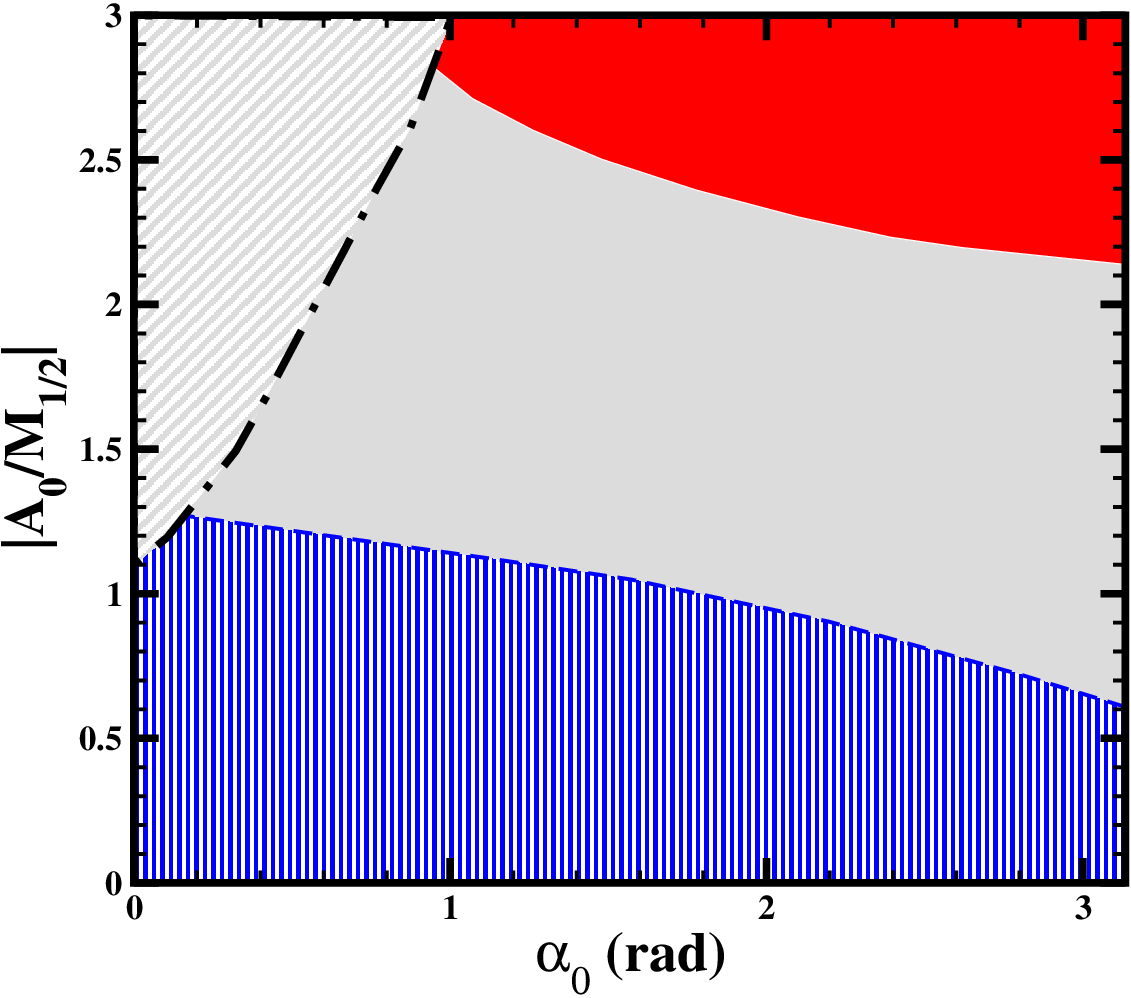}
\caption{Analysis of the neutralino relic density in the 
$|A_0/m_{\frac{1}{2}}|-\alpha_0$ plane for $\xi_3=1.7$~rad, 
$\theta_\mu=0.5$~rad while the other parameters are
the same as in Fig.\ref{xi3thetamu_40}. The area marking 
are the same as in Fig.\ref{a0pha_3180}.
The dark hatched area, contoured by the dashed, line 
predicts $\Omega h^2$ inside the WMAP bounds.
}
\label{a0pha_6080}
\end{figure}

The fluctuation of $\Delta m_b$ with the phases enters in the 
prediction of $m_b(M_Z)$ which is allowed to vary in its 
experimental range. Thus in contrast to the analysis 
of Ref.~\cite{Gomez:2004ek} $h_b$ is not adjusted as the 
phases vary in the present analysis. The value of $h_b$ is 
approximately fixed by the condition $h_b=h_\tau$ at the GUT scale. 
Consequently, the phases do not have a big effect on 
the $\Omega h^2$ prediction in the present scenario. 
For example, in Fig.~(\ref{xi3thetamu_40}), 
$\Omega h^2$ varies only in the approximate range 0.10 -- 0.13. 
The effects of variations with $\alpha_0$ for the point 
Eq.~(\ref{point40}) are analyzed in Fig.~(\ref{a0pha_6080}). 
Specifically, Fig.~(\ref{a0pha_6080}) gives an analysis of 
the neutralino relic density in the $|A|_0-\alpha_0$ plane for 
the input of Eq.~(\ref{point40}) along with 
$\theta_\mu=.5, \; \xi_3=1.7$. One finds a considerable 
structure here exhibiting the important effects of 
$\alpha_0$ in this case.
 The relic density remains within the WMAP bounds, 
in the dark hatched area, while the area above the dashed line 
has a relic density below the lower bound.

\section{ WMAP dark matter constraint and full Yukawa unification}
In the above we discussed the satisfaction of the WMAP relic 
density constraints  consistent with the 
BR($b\rightarrow s+\gamma$) and $b-\tau$ unification constraints
within mSUGRA and its extensions including phases. It was seen 
that the loop corrections 
to the b  quark mass (and to the $\tau$ lepton mass) play an 
important role in accomplishing
$b-\tau$ Yukawa unification at the GUT scale consistent 
with the experimental values for 
the $\tau$ lepton and the b quark masses. The values of $\tan\beta$ 
used in the analysis
above were fairly large, lying in the range up to $40-45$.  
 When  $\tan\beta$ exceeds these values
the possibility that full Yukawa unification for the third generation 
holds becomes feasible. Here we investigate this possibility 
in further detail to determine
if WMAP relic density and the BR($b\rightarrow s+\gamma$) constraints can also
be simultaneously satisfied. In the analysis we will allow 
for the dependence on CP phases.

\begin{figure}
\hspace*{-0.6in}
\centering
\includegraphics[width=12cm,height=8cm]{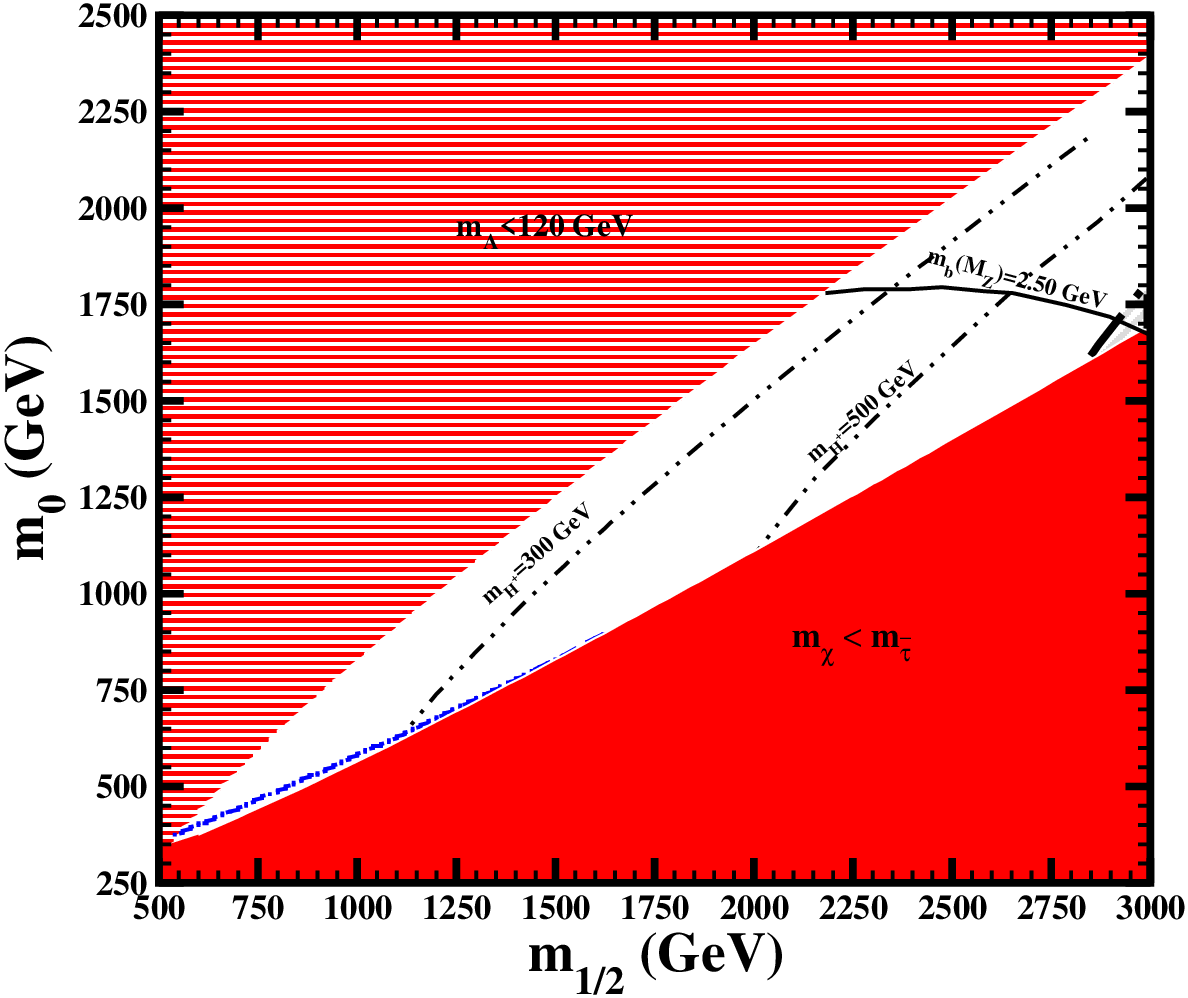}
\caption{Analysis of the neutralino relic density, of 
$BR(b\rightarrow s+\gamma)$ and of $b-t-\tau$ unification 
in mSUGRA when $\mu<0$ ($\theta_{\mu}=\pi$)and $A_0=0$.  
On the upper ruled area $m_A<120$~GeV, while in the 
lower dark shaded area the lightest neutralino is not the LSP. 
The narrow area bounded by dash lines corresponds to 
the WMAP favored relic abundance prediction, while in the area,  
bounded by the dot-dashed thick line, the prediction for 
$BR(b\rightarrow s \gamma$) is inside the experimental bounds. 
The thin dot-dashed line indicates the expansion of 
$BR(b\rightarrow s \gamma$) allowed area when the ratio 
$\frac{m_c}{m_b}=0.29$ is used in computation of the SM contribution. 
For the range of parameters exhibited, the prediction of $m_b(M_Z)$ 
is below the experimental bound, the solid line corresponds to a prediction of $m_b(M_Z)=2.50$~GeV. 
The double-dotted-dashed line corresponds to the indicated 
values of  $m_{H^+}$.}
\label{m0m12_so10}
\end{figure}

In Fig.~(\ref{m0m12_so10}) we present an analysis of full Yukawa 
unification and we also display the constraints of relic density 
and of BR($b\rightarrow s+\gamma$). 
We impose full Yukawa unification at the GUT scale, 
the value of  $tan\beta$ is therefore fixed by the experimental 
$\tau$ and top masses. As before, $m_b(M_Z)$ is a prediction.
Typically, there are two main constraints on $m_0$ and 
$m_{1/2}$ for a  given $A_0$. These are the condition of radiative 
EWSB (or almost equivalently $m_A<120$~GeV) and the condition that 
the LSP be neutral. The constraints on $m_0$ and $m_{1/2}$ such 
that both conditions are met were described in an early 
paper~\cite{shafi} and also emphasized in Ref.\cite{pokorski} 
which gave the relation
\beq
m_A^2=\alpha m_{1/2}^2-\beta m_0^2-{\rm constant}
\eeq
where the coefficients $\alpha$ and $\beta$ are positive and 
$\sim 0.1$, and the constant is $\sim M_Z$. 
Thus for fixed $m_A$ one has a hyperbolic branch. 
Furthermore, the requirement that LSP be neutral, i.e. 
$m_\chi <m_{\tilde{\tau}}$, makes another cut in the allowed area. 
While Fig.~(\ref{m0m12_so10}) exhibits a narrow area
where the WMAP relic density constraint is satisfied, one finds 
that $m_b(M_Z)$ is outside the experimental bounds (the line 
 $m_b(M_Z)=2.50$~GeV is presented as a reference line in Fig.~(\ref{m0m12_so10})).

In the whole figure the value 
of $m_b(M_Z)$ lies below the lower experimental bound. 
The region satisfying the \bbsg bounds is also exhibited in 
Fig.~(\ref{m0m12_so10}). The charged Higgs contribution is 
enhanced in this case due to the low values of its mass, $m_{H^+}$ 
(lines corresponding to the values $m_{H^+}=300$ and $500$~GeV 
are given as reference). Since the SUSY contribution is also positive 
the value of BR($b\rightarrow s+\gamma$) lies below its upper 
experimental limit only for the small region found at 
$m_{1/2}\sim 2900$~GeV. For the SM contribution we followed the 
considerations of~\cite{Gambino:2001ew} by using the 
$\overline{MS}$ running charm mass, so that $\frac{m_c}{m_b}=0.29$.  
In this case the central value of Eq.~\ref{bsg2} for the SM
prediction is obtained. However, as argued in 
Ref.\cite{Hurth,Neubert:2004dd} the theoretical SM 
prediction is possibly lower. Thus as an illustration we also 
give an analysis using the pole mass ratio 
$\frac{m_c}{m_b}=0.29$ which leads to 
a SM prediction of $3.33\times 10^{-4}$.

\begin{figure}
\hspace*{-0.6in}
\centering
\includegraphics[width=12cm,height=8cm]{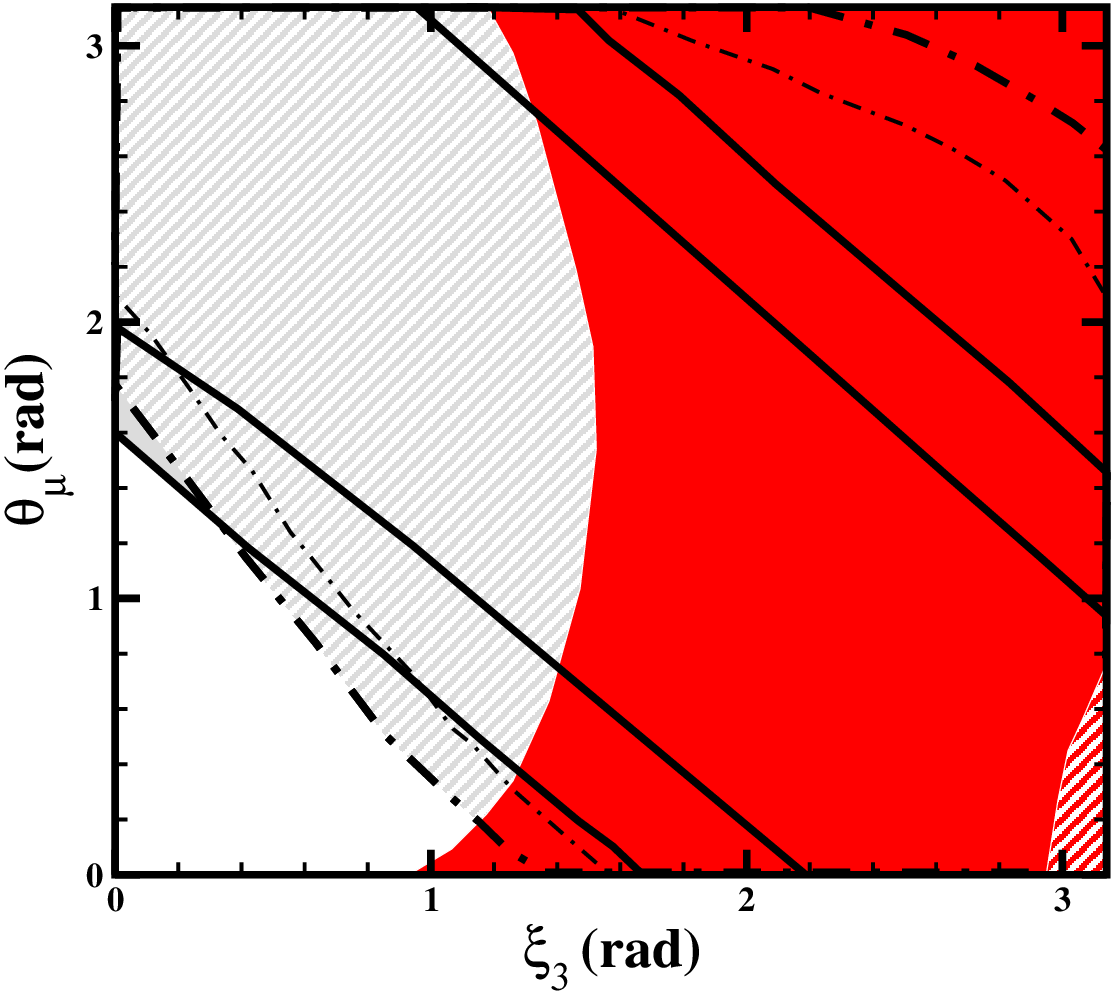}
\caption{The full $b-t-\tau$ unification allowed areas 
(bounded by solid lines) and BR($b\rightarrow s \gamma$) excluded 
areas (bounded by dotted-dashed lines) in the $\theta_\mu-\xi_3$ 
plane, for $m_0=880$~GeV, $M_{1/2}=1500$~GeV, $A_0=0$~GeV 
and all the remaining phases are set to zero. The area within 
the thin dotted-dashed line shows the expansion of the 
$b\rightarrow s \gamma$ allowed area when $\frac{m_c}{m_b}=0.29$. 
In the dark shaded area the lightest neutralino is not the LSP, 
while in the dark hatched area in the right corner $m_A<120$~GeV.} 
\label{xi3thetamu_so10}
\end{figure}

We investigate now the implications of extending the parameter space 
by CP phases for a selected point in the coannihilation region.  
Fig.~(\ref{xi3thetamu_so10}) shows the $\theta_{\mu}-\xi_3$ plane
for $m_0=880$ GeV, $M_{1/2}$=1500 GeV and $A_0=0$. 
The value of $tan\beta$ lies in the range  51 -- 54.5.  
 The prediction for the relic density remains in the 
WMAP range, since the neutralino remains in the coannihilation area.  
The regions where $m_b(M_Z)$ and \bbsg lie inside the 
experimental bounds are shown. There is only 
a rather tiny region, roughly at $\theta_\mu=\pi/2$ and $\xi_3=0$,  
where the Yukawa unification constraints and the 
BR($b\rightarrow s \gamma$) are simultaneously satisfied. 
 This area is affected by the uncertainty in the determination 
of the SM value for BR($b\rightarrow s \gamma$). 
The area is significantly enlarged when the ratio 
$\frac{m_c}{m_b}=0.29$ is used in the BR($b\rightarrow s \gamma$) 
computation. 

\section{Consistency with the EDM constraints.}
With inclusion of phases, one has to account for the satisfaction
of the EDM constraints. In the following we demonstrate that, there 
exist regions in the parameter space, where the WMAP, 
the $b-\tau$ unification, the \bbsg as well as the EDM constraints 
are all satisfied 
when the phases are large. 
In Table 1 we define two points, one for $\tan\beta =40$ 
and another for $\tan\beta =45$ where all constraints are
satisfied as shown in Table 2.

\begin{figure}
\centering
\includegraphics[width=12cm,height=8cm]{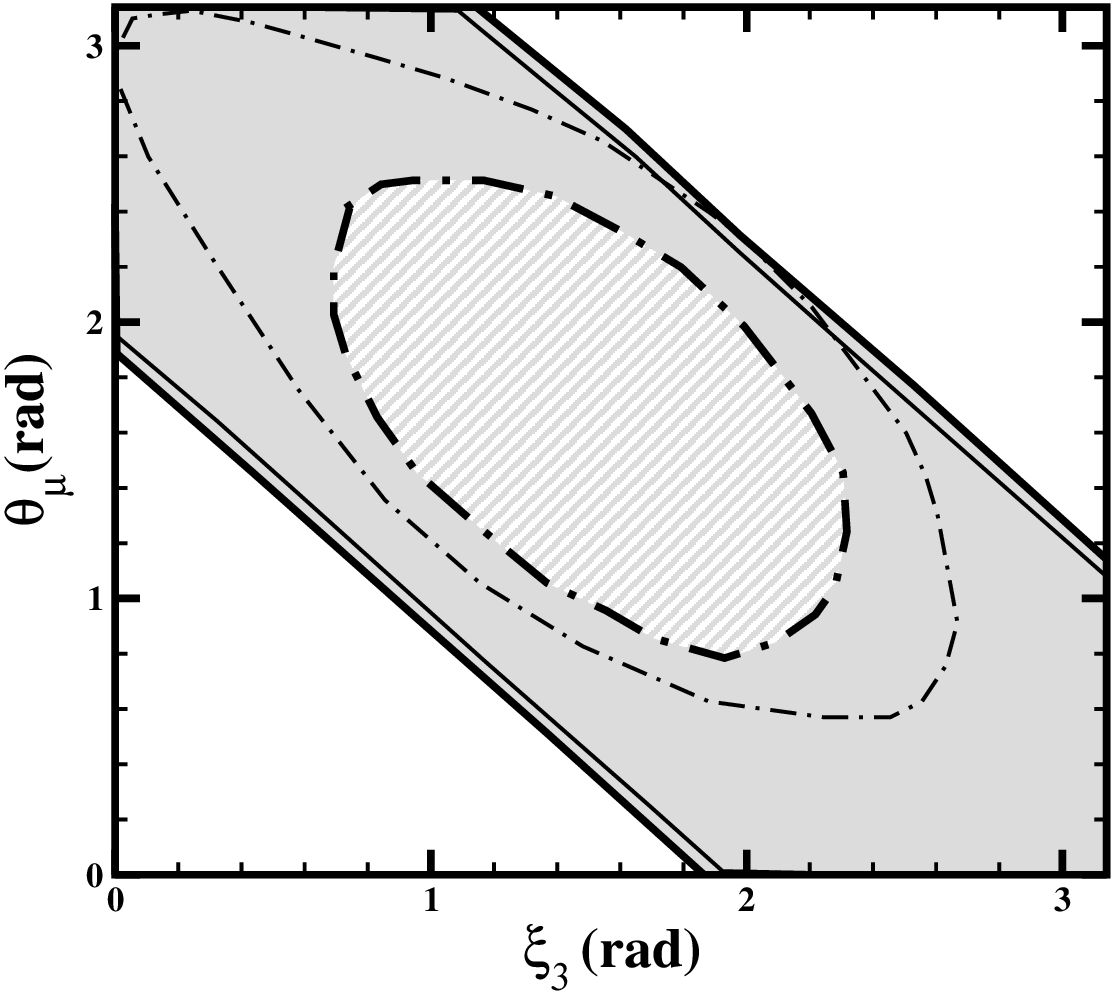}
\caption{Analysis of the $b-\tau$ unification and the 
BR($b\rightarrow s \gamma$) constraints in the 
$\theta_\mu-\xi_3$ plane for $\tan\beta=40$, $\xi_1=1.0$,
$\xi_2=0.15$ and $A_0=0$. $m_0$ and $m_{1/2}$ satisfies 
$m_0=0.832 \cdot m_{1/2}$. In the area contoured by 
solid lines $m_b$ is inside the experimental range, 
while the area inside the dotted-dashed line is 
excluded by the \bbsg bound. The thicker (thiner)  
lines correspond to  $m_{1/2}=1250$~GeV (1050 GeV).}
\label{edm40_1}
\end{figure}

\begin{figure}
\centering
\includegraphics[width=12cm,height=8cm]{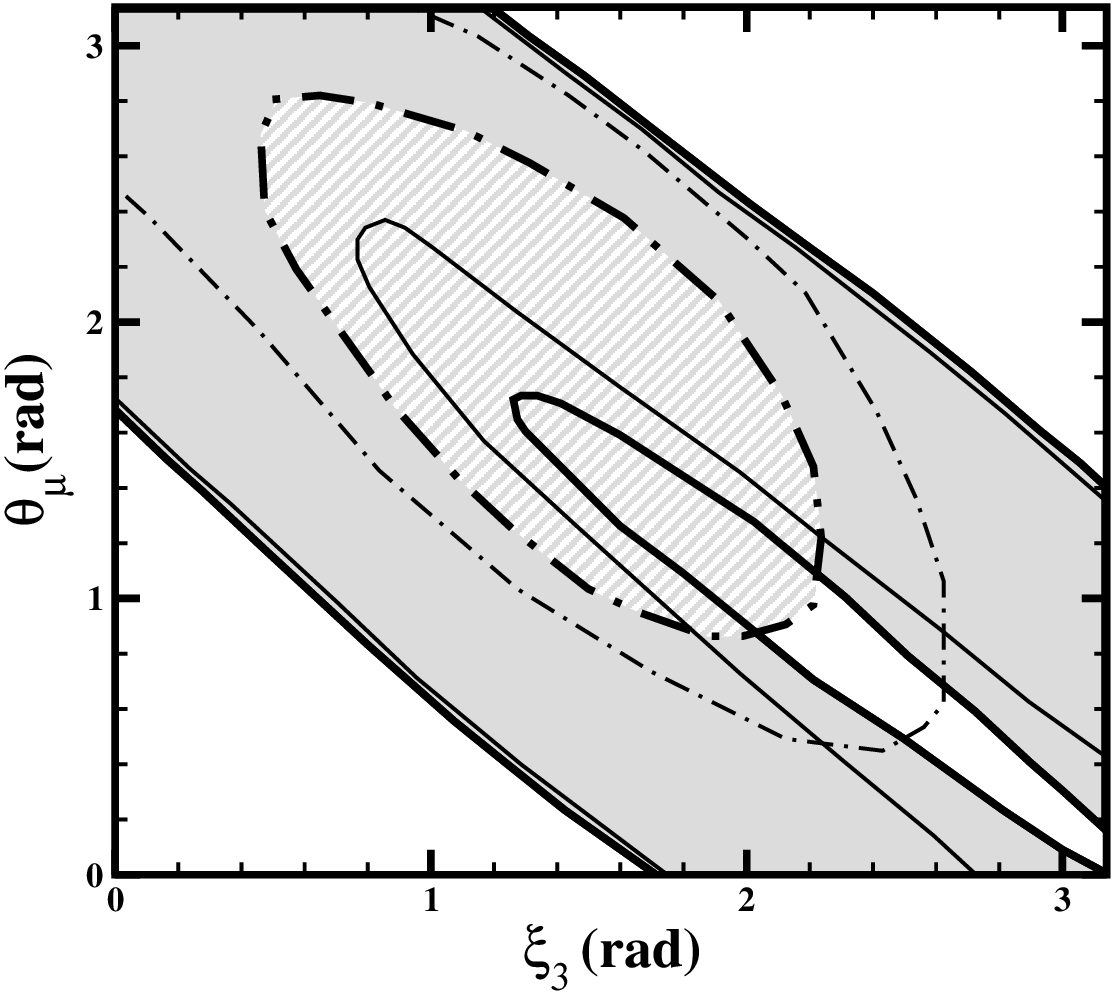}
\caption{Analysis of $b-\tau$ unification and of \bbsg
in the $\theta_\mu-\xi_3$ plane for $\tan\beta=45$, 
$\xi_1=0.5$, $\xi_2=-0.6$ and $A_0=0$.  $m_0$ and $m_{1/2}$
satisfies the equation $m_0=1.8 \cdot m_{1/2}$.  
In the area contoured by solid lines $m_b$ is inside the 
experimental bounds, while the area inside the dot-dash line is 
excluded by the higher(BR($b\rightarrow s \gamma$) bound. 
The thicker (thiner)  
lines correspond to  $m_{1/2}=1000$~GeV (800 GeV).}
\label{edm45_1}
\end{figure}

\begin{figure}
\centering
\includegraphics[width=7.5cm,height=7.5cm]{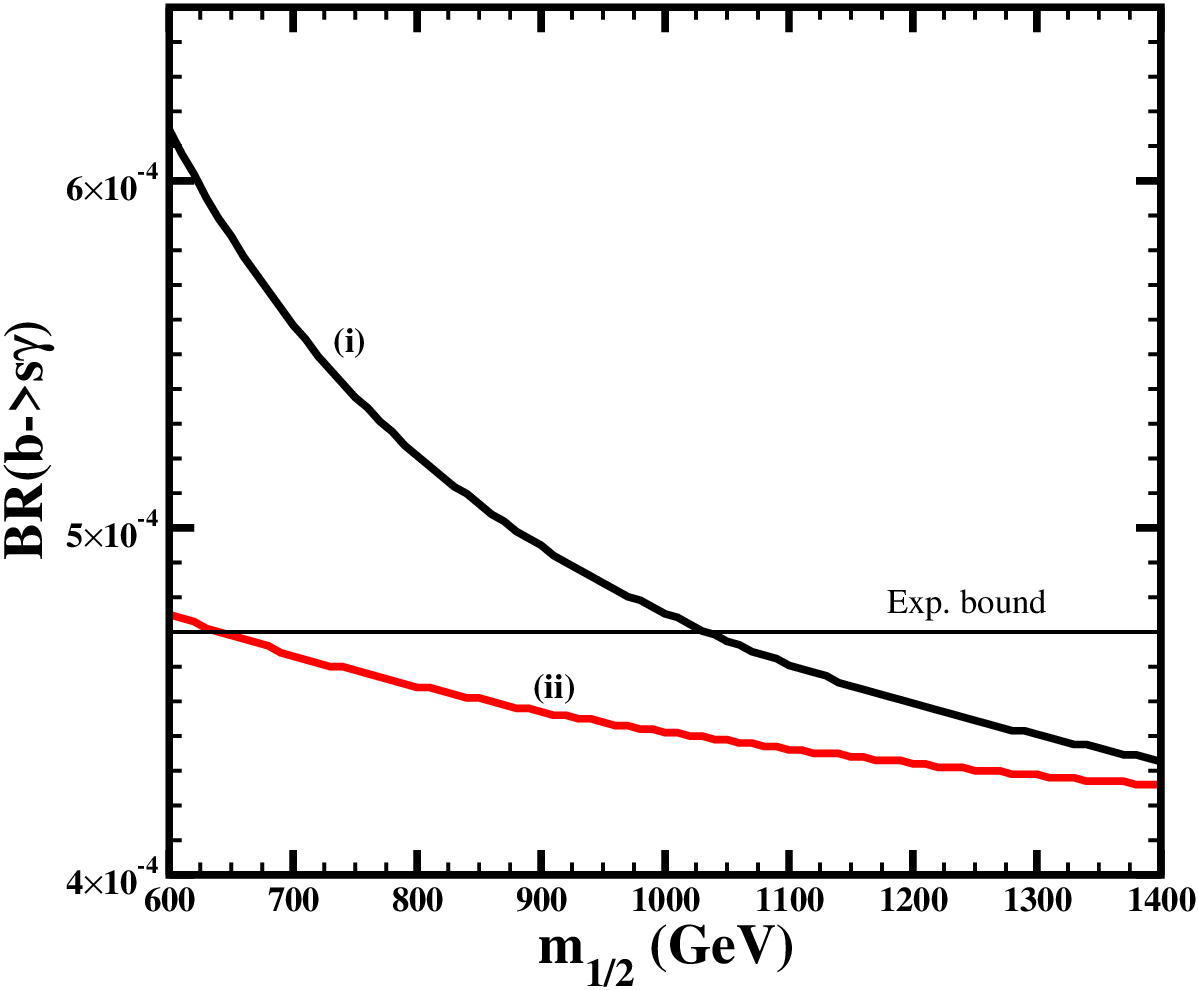}
\includegraphics[width=7.5cm,height=7.5cm]{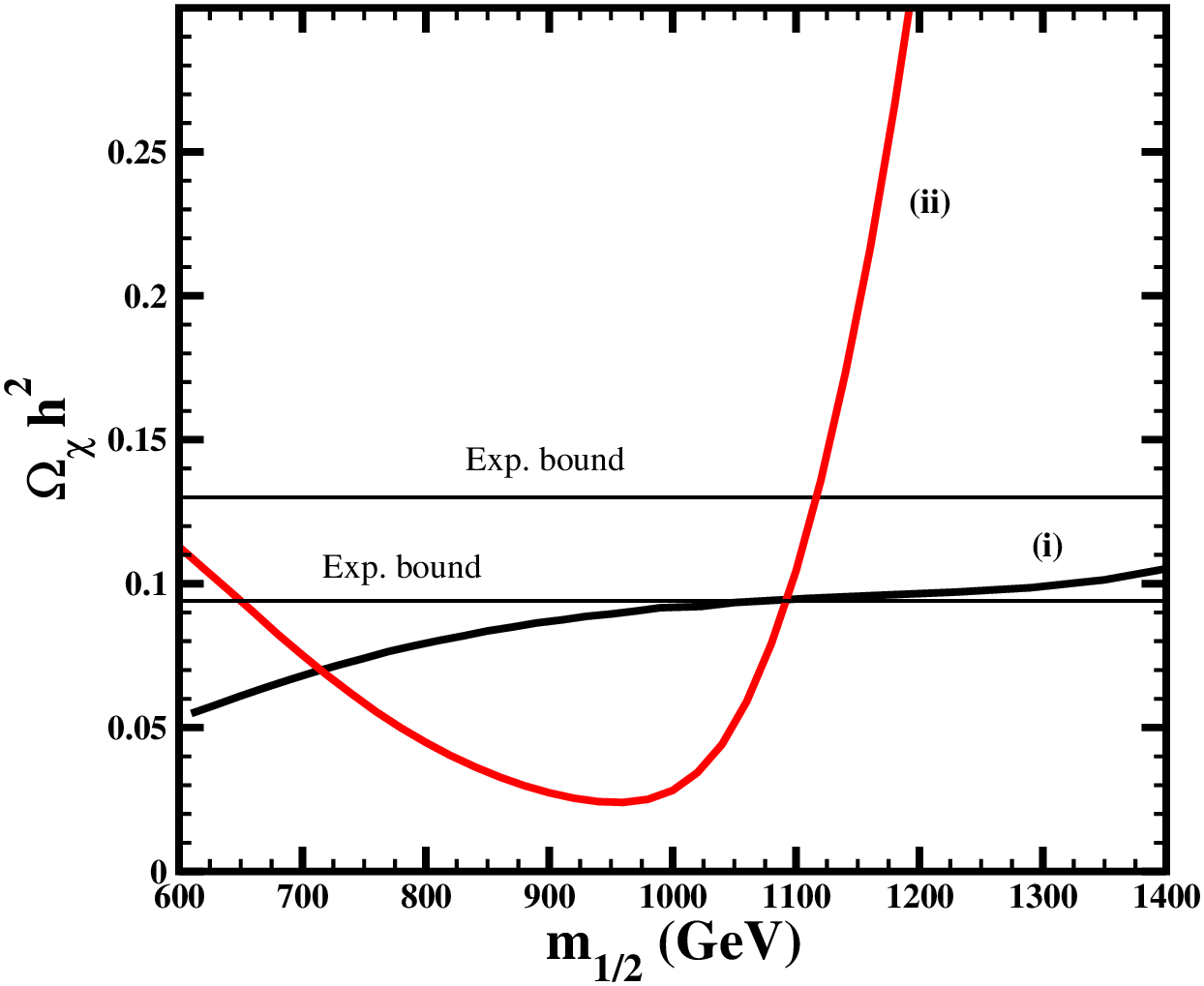}
\caption{Analysis of the relic density and of \bbsg  for 
the points in Table 1. For the case (i), $m_0$ is constrained 
to satisfy $m_0=0.832 \cdot m_{1/2}$ and in the case 
(ii), $m_0=1.8 \cdot m_{1/2}$.}
\label{ohm12}
\end{figure}

A more detailed exhibition of the value of $m_b(M_Z)$ and 
BR($b\rightarrow s+\gamma$) as $\theta_{\mu}$ and $\xi_3$ varies is 
given in Figs.~(\ref{edm40_1}) and ~(\ref{edm45_1}) while  $\xi_1$, $\xi_2$ 
are set at the values given in Table 1.  
It was shown in Ref.\cite{inscaling}, that if the EDM constraints
are satisfied for a given point, there exists a scaling region, 
where $m_0, m_{1/2}, A_0$ scale by a common factor $\lambda$, 
in which the EDM constraints also are satisfied, 
for a reasonable range of $\lambda$ around one.  
The allowed range of $\lambda$ depends on other dynamical parameters. 
For Point (i) $m_0$ and $m_{1/2}$ are related by $m_0= 0.832\cdot m_{1/2}$ 
while for Point (ii) this relation becomes $m_0= 1.80\cdot m_{1/2}$. 
For point (ii) the EDM constrains are satisfied down to 
$m_{1/2}\sim 750$ GeV. 
In  Fig.~(\ref{ohm12}) we display \bbsg and  the neutralino relic 
density for the case of two points in Table 1. 
The analysis shows that 
$m_b(M_Z)$ remains inside its experimental range 
for the range of parameters shown in the figure. 
 The qualitative behavior of the relic density in both cases
can be understood by comparison with the corresponding cases in 
Fig.~\ref{mom12_su5}. For $\tan\beta=40$ the line 
$m_0= 0.832\cdot m_{1/2}$ has a sizeable overlap with the WMAP 
area, whereas for $\tan\beta=45$ the line $m_0= 1.832\cdot m_{1/2}$ 
intersects the WMAP area. The values of $m_b(M_Z)$ ranges 
from 2.80 to 2.86 GeV for the case (i) and from 2.84 to 2.96 GeV 
for case (ii).

\begin{table}[h]
\begin{center}
\begin{tabular}{|r|r|r|r|r|r|r|r|r|r|}
\hline
Point & $m_0$ & $\mhalf$ & $|A_0|$ & $\tan\beta$ & $\theta_\mu$ 
      & $\alpha_A$ & $\xi_1$ & $\xi_2$ & $\xi_3$   \\ \hline
(i)  & 1040 & 1250 & 0 & 40 & 2.9 & 0 & 1.0 & 0.15 & 0.5  \\ \hline
(ii) & 1980 & 1100  & 0 & 45 & 0.6 & 0 & 0.5 & -0.6 & 1.6   \\ \hline
\end{tabular}
\end{center}
\caption{Values of the parameters for point (i) and point (ii).}
\label{points}
\end{table}

\begin{table}[h]
\begin{center}
\begin{tabular}{|r|r|r|r|r|r|r|}
\hline
Point & $|d_e|$ e.cm &  $|d_n|$ e.cm &  $C_{Hg}$ cm & $\Omega h^2$ & 
\bbsg & $m_b(M_Z)$ \\ \hline
(i) & $1.33 \times 10^{-27}$ & $8.87 \times 10^{-27}$ & $1.72 \times 10^{-26}$ 
    & 0.099   & 4.44 $\times 10^{-4}$  & 2.85  \\ \hline 
(ii) & $1.87 \times 10^{-28}$ & $2.71 \times 10^{-26}$ & $1.13 \times 10^{-26}$
    & 0.112   & 4.37$\times 10^{-4}$  & 2.92  \\ \hline 
\end{tabular}
\end{center}
\caption{ The EDMs, the relic abundance $\Omega h^2$, the branching ratio
 \bbsg, and the $m_b(M_Z)$ prediction for point (i) and point (ii) 
as defined in Table \ref{points}.} 
\label{edm_value}
\end{table}

\section{Conclusions}
The main focus of this work is an analysis of the neutralino relic density
consistent with the WMAP data under the constraint of $b-\tau$ Yukawa 
unification, and the constraint of $b\rightarrow s+\gamma$ branching ratio.
 In the analysis of the $b\rightarrow s+\gamma$ branching ratio.
we have included the $\tan\beta$ enhanced NLO corrections which contribute
to the Wilson co-efficients. These enhancements are codified via the
epsilon terms defined in Eq.(13). There is ambiguity in the sign of
some of the terms among the various groups. To resolve this we carried
out an independent calculation of these quantities as discussed in Sec.3.
The analysis is carried out within SUGRA unified models where 
universality on the magnitudes of soft parameters 
at the GUT scale is assumed, but we allow for CP 
violating phases and specifically allow
non-universality of the phases in the gaugino mass sector. 
First we give an analysis for the case when all the 
soft parameters are real. This is the mSUGRA case, and here
we find that for values of $\tan\beta$ in the range 27-48, 
one obtains an amount of dark matter consistent with WMAP 
as well as consistency with $b-\tau$ unification and with 
the $b\rightarrow s+\gamma$ constraint. An interesting 
phenomenon that arises is the following: There are three 
regions in the $m_0-m_{\frac{1}{2}}$ parameter space where
relic density and the  $b\rightarrow s+\gamma$ constraint 
can be satisfied in general. These consist of the 
coannihilation region, the resonance region, and the HB/FP region.
Of these only the first two can satisfy the Yukawa unification 
constraint. Thus the constraint of Yukawa unification narrows 
the available parameter space by eliminating the HB/FP region. 
We then extend this analysis to include phases and show that 
new regions of the parameter space allow for consistency with 
the WMAP data and other constraints extending the allowed 
region of the parameter space. 
      In b-$\tau$ unification case we find explicit phase
      arrangements such that the EDM bounds are satisfied,
      $m_b(M_Z)$ and the rate for \bbsg  lie within their
      experimental ranges, and  the prediction of the neutralino
      relic density lies within the WMAP bounds.
 We  have also 
given an analysis of the full $b-\tau -t$ Yukawa unification constraint with 
inclusion of CP phases. We find a small area where $m_b$ is predicted 
inside the experimental range and the \bbsg bound is satisfied. 
Furthermore, the relic density of neutralinos lies within the 
WMAP bounds due to $\chi-\tilde{\tau}$ coannihilations. 
However, this area is rather small and moreover we could not find 
phase arrangements satisfying the EDM constraints. 
It is conjectured that inclusion of additional non-universalities 
could rectify the situation.

\vspace{1cm}
\noindent
{\large\bf Acknowledgments}\\ 
MEG acknowledges support from the 'Consejer\'{\i}a de Educaci\'on de 
la Junta de Andaluc\'{\i}a', the Spanish DGICYT under contract 
BFM2003-01266 and European Network for Theoretical Astroparticle 
Physics (ENTApP), member of ILIAS, EC contract number
RII-CT-2004-506222. 
The research of TI and PN was supported in part by NSF grant PHY-0139967. 
PN also acknowledges support from the Alexander von Humboldt Foundation and
thanks the Max Planck Institute, Munich for hospitality extended him.
SS is supported by Fundac\~{a}o de Amparo \`{a} Pesquisa do
Estado de S\~{a}o Paulo (FAPESP).

\newpage

\end{document}